%% file: TF_180526.tex
%
\newif\ifloadreferences\loadreferencestrue
%
\input preamble %
%
%
%
%
%
\def\Pagetitle{\hfil}
\def\Pagefooter{\hfil}
\font\tablefont=cmr7
\newif\ifshowaddress\showaddresstrue
\null \vfill
\def\centre{\rightskip=0pt plus 1fil \leftskip=0pt plus 1fil \spaceskip=.3333em \xspaceskip=.5em \parfillskip=0em \parindent=0em}%
\def\textmonth#1{\ifcase#1\or January\or Febuary\or March\or April\or May\or June\or July\or August\or September\or October\or November\or December\fi}
\font\abstracttitlefont=cmr10 at 14pt {\abstracttitlefont\centre Sur un syst\`eme int\'egrable \`a bord.\par}
\bigskip
{\centre 8 juin 2018\par}
\bigskip
{\centre Martin Kilian\footnote{${}^1$}{{\tablefont School of Mathematical Sciences, Western Gateway Building, University College, Cork, WGB1.61, Ireland\hfill}},
Graham Smith\footnote{${}^2$}{{\tablefont Instituto de Matem\'atica, UFRJ, Av. Athos da Silveira Ramos 149, Centro de Tecnologia - Bloco C, Cidade Universit\'aria - Ilha do Fund\~ao, Caixa Postal 68530, 21941-909, Rio de Janeiro, RJ - BRAZIL\hfill}}\par}
\bigskip
\noindent{\bf R\'esum\'e~:~}Nous d\'eveloppons de nouvelles applications du formalisme de $K$-matrice de Sklyanin \`a l'\'etude de solutions p\'eriodiques de l'\'equation sinh-Gordon.
\bigskip
\noindent{\bf Classification AMS~:~}37K10
%
%
\par
\vfill
\eject
%
\global\pageno=1
\myfontdefault
\def\Pagetitle{\hfil Sur un syst\`eme int\'egrable \`a bord.\hfil}
\def\Pagefooter{\hfil{\myfontdefault\folio}\hfil}
\catcode`\@=11
\def\multiline#1{\null\,\vcenter{\openup1\jot \m@th %
\ialign{\strut$\displaystyle{##}$\hfil\crcr#1\crcr}}\,}
\def\eqalignleft#1{\null\,\vcenter{\openup1\jot \m@th %
\ialign{\strut$\displaystyle{##}$\hfil&$\displaystyle{{}##}$\hfil\crcr#1\crcr}}\,}
\def\triplealign#1{\null\,\vcenter{\openup1\jot \m@th %
\ialign{\strut\hfil$\displaystyle{##}\quad$&\hfil$\displaystyle{{}##}$&$\displaystyle{{}##}$\hfil\crcr#1\crcr}}\,}
\def\odeICalign#1{\null\,\vcenter{\openup1\jot \m@th %
\ialign{\strut\hfil$\displaystyle{##}$&$\displaystyle{{}##}$
\hfil\quad&\hfil$\displaystyle{##}$&$\displaystyle{{}##}$\hfil&
\hfil\quad$\displaystyle{##}$&$\displaystyle{{}##}$\hfil\crcr#1\crcr}}\,}
\def\tripleeqalign#1{\null\,\vcenter{\openup1\jot \m@th %
\ialign{\strut$\displaystyle{##}\quad$\hfil&$\displaystyle{{}##}$\hfil&$\displaystyle{{}##}$\hfil\crcr#1\crcr}}\,}
\catcode`\@=12

\makeop{sl}
\makeop{T}
\def\proof{{\noindent\bf D\'emonstration~:\ }}%
\def\remark{{\noindent\bf Remarque~:\ }}%
\newref{AdlerEtAl}{Adler V., G\"urel B., G\"urses M., Habublin I., Boundary conditions for integrable equations, {\sl J. Phys. A Math. Gen.}, {\bf 30}, (1997), 3505--3513}
\newref{BajnokEtAl}{Bajnok Z., Palla L., Tak\'acs G., Boundary sine-Gordon model, arXiv:hep-th/0211132}
\newref{BobenkoBook}{Belokolos A. D., Bobenko A. I., Enol'skii V. Z., Its A. R., Matveev V. B., {\sl Algebro-geometric approach to nonlinear integrable equations}, Springer-Verlag, (1994)}
\newref{Bobenko}{Bobenko A. I., All constant mean curvature tori in $\Bbb{R}^3$, $\Bbb{S}^3$, $\Bbb{H}^3$ in terms of theta-functions, {\sl Math. Ann.}, {\bf 290}, (1991), 209--245}
\newref{BobenkoKuksin}{Bobenko A. I., Kuksin S. B., Small amplitude solutions of the sine-Gordon equation on an interval under Dirichlet or Neumann boundary conditions, {\sl J. Nonlinear Sci.}, {\bf 5}, (1995), 207--232}
\newref{BurstallFerusPeditPinkall}{Burstall F. E., Ferus D., Pedit F., Pinkall U., Harmonic tori in symmetric spaces and commuting hamiltonian systems on loop algebras, {\sl Ann. of Math.}, {\bf 138}, no. 1, (1993), 173--212}
\newref{CorriganA}{Corrigan E., Recent developments in affine Toda quantum field theory, in {\sl Particles and Fields}, CRM Series in Mathematical Physics, Springer, New York, NY, (1991)}
\newref{CorriganEtAlA}{Corrigan E., Dorey P. E., Rietdijk R. H., Sasaki R., Affine Toda field theory on a half line, {\sl Phys. Lett.} {\bf B}, {\bf 333}, (1994)}
\newref{CorriganEtAlB}{Corrigan E., Dorey P. E., Rietdijk R. H., Aspects of affine Toda field theory on a half line, {\sl Prog. Theor. Phys. Suppl.}, {\bf 118}, (1995), 143--164}
\newref{Delius}{Delius G. W., Soliton preserving boundary condition in affine Toda field theories, {\sl Phys. Lett.} {\bf B}, {\bf 444}, (1998), 217--223}
\newref{FadeevTakhtajan}{Fadeev L. D., Takhtajan L., {\sl Hamiltonian methods in the theory of solitons}, Classics in Mathematics, Spring-Verlag, Berling, Heidelberg, (2007)}
\newref{FokasIts}{Fokas A. S., Its. A. R., An initial value problem for the sine-Gordon equation in laboratory coordinates, {\sl Theor. Math. Phys.}, {\bf 92}, no. 3, 964--978}
\newref{Fokas}{Fokas A. S., Linearizable initial value problems for the sine-Gordon equation on the half-line, {\sl Nonlinearity}, {\bf 17}, (2004), 1521--1534}
\newref{FordyWood}{Fordy A. P., Wood J. C., {\sl Harmonic maps and integrable systems}, Aspects of Mathematics, {\bf E23}, Vieweg, Braunschweig, (1994)}
\newref{GilbTrud}{Gilbarg D., Trudinger N. S., {\sl Elliptic partial differential equations of second order}, Classics in Mathematics, Springer-Verlag, Berlin, (2001)}
\newref{HauswirthKillianSchmidt}{Hauswirth L., Killian M., Schmidt M., Finite type minimal annuli in $\Bbb{S}^2\times\Bbb{R}$, {\sl Illinois J. Math.}, {\bf 57}, no. 3, (2013), 697--741}
\newref{Hitchin}{Hitchin N., Harmonic maps from a $2$-torus to the $3$-sphere, {\sl J. Diff. Geom.}, {\bf 31}, no. 3, (1990), 627--710}
\newref{MacIntyre}{MacIntyre A., Integrable boundary conditions for classical sine-Gordon theory, {\sl J. Phys. A. Math. Gen.}, {\bf 28}, (1995), 1089--1100}
\newref{PinkallSterling}{Pinkall U., Sterling I, On the Classification of Constant Mean Curvature Tori, {\sl Ann. of Math.}, {\bf 130}, no. 2, (1989), 407--451}
\newref{SklyaninI}{Sklyanin E. K., Boundary conditions for integrable equations, {\sl Funct. Anal. Appl.}, {\bf 21}, (1987), 164--166}
\newref{SklyaninII}{Sklyanin E. K., Boundary conditions for integrable quantum systems, {\sl J. Phys. A. Math. Gen.}, {\bf 21}, (1988), 2375--2389}
%
%
\makeop{D}
\newsubhead{Introduction}[Introduction]
La th\'eorie de Toda affine associe \`a chaque alg\`ebre de Lie semi-simple complexe $\frak{g}$ un unique syst\`eme int\'egrable d'\'equations aux d\'eriv\'ees partielles (c.f. \cite{CorriganA} \& \cite{Delius}). Pour un ouvert $\Omega$ de $\Bbb{C}$ et une fonction lisse $\omega:\Omega\rightarrow\frak{g}^*$, ce syst\`eme s'\'ecrit
$$
\omega_{z\overline{z}} + m^2\sum_{k=0}^r n_k\alpha_k e^{\langle\alpha_k,\omega\rangle}=0,\eqnum{\nexteqnno[GeneralTodaSystem]}
$$
o\`u $r$ est le rang de $\frak{g}$, $\langle\cdot,\cdot\rangle$ sa forme de Killing, $m$ un param\`etre de masse, $\alpha_1,\cdots,\alpha_r$ un syst\`eme de racines de $\frak{g}$, $n_1,\cdots,n_r$ des coefficients entiers d\'etermin\'es par $\frak{g}$, et
$$
n_0\alpha_0 := -\sum_{i=1}^r n_i\alpha_i.\eqnum{\nexteqnno[DefinitionOfAlphaZero]}
$$
Dans le cas particulier o\`u $\frak{g}=\opsl(2,\Bbb{C})$, nous r\'ecuperons l'\'equation sinh-Gordon
$$
\omega_{z\overline{z}} + \frac{1}{8}\opSinh(2\omega) = 0,\eqnum{\nexteqnno[SinhGordon]}
$$
qui est le plus simple de tous les syst\`emes de Toda.
\par
L'\'equation sinh-Gordon porte d\'ej\`a en elle une \'etonnante richesse de structures alg\'e\-briques. En effet (c.f. \cite{FadeevTakhtajan}), en le concevant heuristiquement comme une famille bidimensionelle involutive de flots hamiltoniens sur $\opT^*C^\infty(\Bbb{R})$, nous voyons que l'int\'egrabilit\'e de cette \'equation se manifeste en l'existence d'un syst\`eme de coordonn\'ees ``action-angle'' de cet espace. Il existe de plus une suite croissante
$$
X_1\subseteq X_2\subseteq\cdots
$$
de sous-vari\'et\'es de dimension finie de $\opT^*C^\infty(\Bbb{R})$ qui sont toutes pr\'eserv\'ees par ces flots-ci. Les solutions, dites de {\sl type fini}, qui appartiennent \`a la r\'eunion de ces sous-vari\'et\'es sont donn\'ees par des formules explicites relativement \'el\'ementaires (c.f. \cite{BobenkoBook} \& \cite{Bobenko}) et sont alors d'un int\'er\^et particulier dans la th\'eorie de l'\'equation sinh-Gordon.
\par
Le probl\`eme d'\'etablir sous quelles conditions une solution donn\'ee est de type fini est alors un enjeu important de l'\'etude de l'\'equation sinh-Gordon. Dans la suite, nous allons aborder ce probl\`eme du point de vu de la th\'eorie Adler-Kostant-Symes (c.f. \cite{BurstallFerusPeditPinkall} et \cite{FordyWood}). Rappelons d'abord que l'\'equation sinh-Gordon se traduit en la condition d'int\'egrabilit\'e de la paire de Lax
$$\eqalign{
\alpha_z(\lambda,\gamma) &= \frac{1}{2}\omega_z\sigma_0 + \frac{i}{4\lambda}e^\omega\sigma_+ + \frac{i\gamma}{4}e^{-\omega}\sigma_-\ \text{et}\cr
\alpha_{\overline{z}}(\lambda,\gamma) &= -\frac{1}{2}\omega_{\overline{z}}\sigma_0 + \frac{i}{4\gamma}e^{-\omega}\sigma_+ + \frac{i\lambda}{4}e^{\omega}\sigma_-,\cr}\eqnum{\nexteqnno[LaxPair]}
$$
o\`u
$$
\sigma_0 := \pmatrix 1 & 0\cr 0& -1\cr\endpmatrix,\
\sigma_+ := \pmatrix 0 & 1\cr 0& 0\cr\endpmatrix\ \text{et}\
\sigma_- := \pmatrix 0 & 0\cr 1& 0\cr\endpmatrix,\eqnum{\nexteqnno[SigmaMatricesI]}
$$
et $\lambda,\gamma\in S^1$ sont des param\`etres complexes {\sl unitaires} que nous appelons respectivement {\sl param\`etre spectral} et {\sl param\`etre de torsion}.\footnote{${}^*$}{La paire de Lax appara\^\i t d\'ej\`a sous cette forme de mani\`ere implicite dans le travail \cite{PinkallSterling} de Pinkall \& Sterling. En effet, alors que le param\`etre de torsion y appara\^\i t explicitement, le param\`etre spectral appara\^\i t pourtant implicitement dans les expansions des fonctions g\'en\'eratrices des suites qui y sont \'etudi\'ees.} Nous appelons {\sl champ de Killing}\footnote{${}^\ddagger$}{En fait, nous allons utiliser dans la suite une d\'efinition un peu plus pr\'ecise des champs de Killing (c.f. \eqnref{KillingFieldEquation} ci-dessous).} de $\omega$ toute fonction $\Phi:S^1\times S^1\rightarrow C^\infty(\Omega,\opsl_2(\Bbb{C}))$ solution du syst\`eme d'\'equations \`a d\'eriv\'ees partielles
$$
d\Phi(\lambda,\gamma) = [\Phi(\lambda,\gamma),\alpha(\lambda,\gamma)],\eqnum{\nexteqnno[KillingFieldEqnIntro]}
$$
et nous disons qu'un champ de Killing est {\sl polyn\^omial} lorsqu'il s'\'ecrit sous la forme
$$
\Phi(\lambda) = \sum_{(m,n)\in A}^k\Phi_{m,n}\lambda^m\gamma^n,\eqnum{\nexteqnno[PolynomialKillingFieldIntro]}
$$
o\`u $A$ est un sous-ensemble {\sl fini} de $\Bbb{Z}\times\Bbb{Z}$ et, pour tout $(m,n)$, $\Phi_{m,n}:\Omega\rightarrow\opsl_2(\Bbb{C})$ est une fonction lisse. La th\'eorie Adler-Kostant-Symes montre alors qu'une solution est de type fini si et seulement si elle admet un champ de Killing polyn\^omial. En particulier, en suivant ce genre de raisonnement, Pinkall \& Sterling montrent dans \cite{PinkallSterling} que toute solution sur le tore - c'est-\`a-dire, toute solution doublement p\'eriodique sur $\Bbb{C}$ - est de type fini, ce qui permet \`a Bobenko d'obtenir dans \cite{Bobenko} des formules explicites pour ces solutions en termes des fonctions theta (c.f. aussi \cite{Hitchin}).
\par
Dans cet article, nous allons nous int\'eresser \`a des solutions de l'\'equation sinh-Gordon sur l'anneau $\Sigma:=S^1\times[-T,T]$. Pour cela, il faut introduire des conditions au bord, dites {\sl int\'egrables}, qui pr\'eservent l'int\'egrabilit\'e du syst\`eme. La th\'eorie classique montre relativement facilement que des conditions au bord de Dirichlet et de Neumann sont int\'egrables (c.f. \cite{BobenkoKuksin}). Pourtant, l'identification de conditions au bord int\'egrables plus g\'en\'erales s'est r\'evel\'ee un probl\`eme non-trivial qui a d\^u attendre sa r\'esolution, de mani\`ere ind\'ependante, dans les travaux de Sklyanin, d'un c\^ot\'e, et du groupe de Durham, de l'autre.\footnote{${}^\dagger$}{Nous r\'ef\'erons le lecteur \`a l'article \cite{MacIntyre} de MacIntyre pour une excellente exposition de ces id\'ees.} En effet, les travaux \cite{CorriganEtAlA} et \cite{CorriganEtAlB} du groupe de Durham montrent l'int\'egrabilit\'e des conditions au bord
$$
\omega_y = Ae^\omega + Be^{-\omega},\eqnum{\nexteqnno[DurhamBoundaryConditions]}
$$
o\`u $A$ et $B$ sont constantes le long de chaque composant du bord, et nous appelons alors ces conditions des {\sl conditions au bord de Durham}. De mani\`ere compl\'ementaire, les travaux \cite{SklyaninI} \& \cite{SklyaninII} de Sklyanin introduisent un formalisme alg\'ebrique qui exprime des conditions au bord int\'egrables en termes de la matrice
$$
K(\lambda,\gamma) := \pmatrix 4A\gamma-4B\lambda&\frac{\lambda}{\gamma} -\frac{\gamma}{\lambda}\cr \frac{\lambda}{\gamma} -\frac{\gamma}{\lambda}&\frac{4A}{\gamma}-\frac{4B}{\lambda}\cr\endpmatrix,\eqnum{\nexteqnno[KMatrixIntro]}
$$
o\`u $A$ et $B$ sont constantes le long de chaque composante de bord de $\Sigma$. Nous appelons cette matrice {\sl matrice de Sklyanin}. Enfin (c.f. Section \subheadref{SklyaninsKMatrix}), une solution $\omega$ de l'\'equation sinh-Gordon v\'erifie les conditions au bord de Durham si et seulement si la partie r\'eelle $\alpha_x$ de sa paire de Lax v\'erifie en tout point de $\partial\Sigma$ la relation
$$
K(\lambda,\gamma)\alpha_x(\lambda,\gamma) = \alpha_x\bigg(\frac{1}{\lambda},\frac{1}{\gamma}\bigg)K(\lambda,\gamma),\eqnum{\nexteqnno[SklyaninConditionLaxIntro]}
$$
pour tous $\lambda,\gamma\in S^1$, et nous appelons cette condition {\sl condition de Sklyanin pour des paires de Lax}.
\par
Nous allons montrer alors deux choses~: nous allons montrer d'abord comment les conditions au bord de Durham se traduisent dans le contexte de la th\'eorie Adler-Kostant-Symes, et nous allons montrer ensuite comment en d\'eduire que les solutions sont de type fini. En effet, nous disons qu'un champ de Killing $\Phi$ v\'erifie la {\sl condition de Sklyanin pour des champs} lorsque, en tout point de $\partial\Sigma$,
$$
K(\lambda,\gamma)\Phi(\lambda,\gamma) = \overline{\Phi\bigg(\frac{1}{\lambda},\frac{1}{\gamma}\bigg)}^t K(\lambda,\gamma),\eqnum{\nexteqnno[SklyaninConditionIntro]}
$$
pour tous $\lambda,\gamma\in S^1$.
\proclaim{Theorem \nextprocno}
\noindent Soit $\Sigma:=S^1\times[-T,T]$. Soit $\omega:\Sigma\rightarrow\Bbb{R}$ solution de l'\'equation sinh-Gordon. Alors $\omega$ v\'erifie les conditions au bord de Durham si et seulement si elle admet un champ de Killing polynomial $\Phi$ qui v\'erifie la condition de Sklyanin pour des champs. En particulier, toute solution $\omega$ de l'\'equation sinh-Gordon avec des conditions au bord de Durham est de type fini.
\endproclaim
\proclabel{MainTheorem}
{\bf\noindent Remarque~:\ }Nous d\'emontrons le th\'eor\`eme \procref{MainTheorem} dans la section \subheadref{DesChampsDeKillingPolynomes}.
\medskip
{\bf\noindent Remerciements~:\ }Enfin, nous remercions chaleureusement Laurent Hauswirth pour des \'echanges en\-rich\-issants et des commentaires importants sur des versions ant\'erieures de cet article.
\newsubhead{Les champs de Killing}[LesChampsDeKilling]
Nous introduisons d'abord un nouveau formalisme qui \'eclaircit les structures alg\'ebriques que nous allons \'etudier. Dans tout ce qui suit, nous allons traiter $\lambda$ comme une variable et $\gamma$ comme une constante. De plus, dans un premier temps, nous n'allons \'etudier que le cas o\`u $\gamma=1$. En effet, pour tous $\lambda,\gamma\in S^1$, nous avons
$$
\alpha(\lambda,\gamma) = e^{-\frac{\theta}{2}\sigma_0}\alpha\bigg(\frac{\lambda}{\gamma},1\bigg)e^{\frac{\theta}{2}\sigma_0},\eqnum{\nexteqnno[GaugeTransform]}
$$
o\`u $\theta\in\Bbb{R}$ v\'erifie
$$
e^{2i\theta}=\gamma,
$$
et le cas g\'en\'eral s'obtient alors en appliquant la transformation de jauge \eqnref{GaugeTransform}.
\par
Soit $X$ une vari\'et\'e et soit $E$ un espace vectoriel complexe. Pour $k\in\Bbb{Z}$, appelons {\sl s\'erie de Laurent} de degr\'e $k$ sur $X$ \`a valeurs dans $E$ toute s\'erie formelle de la forme
$$
\Phi(\lambda) := \sum_{m=k}^\infty\Phi_m\lambda^m,
$$
o\`u, pour tout $m$, $\Phi_m$ est une fonction lisse sur $X$ \`a valeurs dans $E$ et $\Phi_k$ est diff\'erente de z\'ero. Appelons aussi la s\'erie triviale
$$
\Phi(\lambda) = 0
$$
{\sl s\'erie de Laurent} de degr\'e $+\infty$. Notons $\Cal{L}(X,E)$ l'espace de s\'eries de Laurent sur $X$ \`a valeurs dans $E$. Notons aussi
$$\eqalign{
\Cal{L}(X) &:= \Cal{L}(X,\Bbb{C})\ \text{et}\cr
\Cal{L} &:= \Cal{L}(\left\{x\right\},\Bbb{C}),\cr}\eqnum{\nexteqnno[SpecialSpacesOfLaurentSeries]}
$$
o\`u $\left\{x\right\}$ est la vari\'et\'e qui consiste en un seul point. Nous v\'erifions le lemme \'el\'ementaire suivant~:
\proclaim{Lemme \nextprocno}
\myitem{(1)} $\Cal{L}(X,E)$ est un espace vectorial complexe;
\medskip
\myitem{(2)} $\Cal{L}(X)$ est une alg\`ebre commutative et unitaire;
\medskip
\myitem{(3)} $\Cal{L}$ est un corps; et
\medskip
\myitem{(4)} $\Cal{L}(X,E)$ est un module sur $\Cal{L}(X)$ et un espace vectoriel sur $\Cal{L}$.
\endproclaim
\proclabel{BasicAlgebraicPropertiesOfL}
\par
Pour $k\leq l\in\Bbb{N}$, appelons {\sl polyn\^ome de Laurent} de bidegr\'e $(k,l)$ sur $X$ toute s\'erie de Laurent de la forme
$$
\Phi(\lambda) := \sum_{m=k}^l\Phi_m\lambda^m,
$$
o\`u $\Phi_k$ et $\Phi_l$ sont diff\'erentes de z\'ero. De nouveau, appelons aussi la somme triviale
$$
\Phi(\lambda) := 0
$$
{\sl polyn\^ome de Laurent} de bidegr\'e $(+\infty,+\infty)$. Notons $\Cal{P}(X,E)$ l'espace de polyn\^omes de Laurent sur $X$ \`a valeurs dans $E$.
\par
%
La paire de Lax \eqnref{LaxPair} est une $1$-forme sur $\Sigma=S^1\times[-T,T]$ \`a valeurs dans $\Cal{P}(\Sigma,\opsl(2,\Bbb{C}))$. Rappelons qu'un {\sl champ de Killing} sur $\Sigma$ est une s\'erie de Laurent $\Phi\in\Cal{L}(\Sigma,\opsl(2,\Bbb{C}))$ qui v\'erifie
$$
d\Phi = [\Phi,\alpha]\eqnum{\nexteqnno[KillingFieldEquation]}
$$
en tout point de $\Sigma$ (c.f. \cite{BurstallFerusPeditPinkall}).\footnote{${}^*$}{Dans \cite{BurstallFerusPeditPinkall}, les auteurs appellent ces objets {\sl champs de Killing formels}. Comme nous trouvons l'adjectif ``formel'' superflu dans le cadre actuel, nous pr\'ef\'erons l'omettre.}
\proclaim{Lemme \nextprocno, (c.f. \cite{HauswirthKillianSchmidt})}
\noindent Soit
$$
\Phi := \sum_{m=k}^\infty\pmatrix u_m& e^\omega t_m\cr e^\omega s_m& -u_m\cr\endpmatrix \lambda^m
$$
une s\'erie de Laurent sur $\Sigma$ \`a valeurs dans $\opsl(2,\Bbb{C})$. Alors $\Phi$ est champ de Killing si et seulement si, en tout point de $\Sigma$ et pour tout $m$,
$$\eqalignno{
4u_{m,z} + i e^{2\omega} s_{m+1} - i\gamma\tau_m &= 0,&\nexteqnno[RI]\cr
4u_{m,\overline{z}} + i\gamma^{-1}s_m - i e^{2\omega}t_{m-1} &= 0,&\nexteqnno[RII]\cr
4\omega_z t_m + 2 t_{m,z} - i u_{m+1} &= 0,&\nexteqnno[RIII]\cr
2e^\omega t_{m,\overline{z}} - i \gamma^{-1}e^{-\omega} u_m &= 0,&\nexteqnno[RIV]\cr
2e^\omega s_{m,z} + i \gamma e^{-\omega} u_m &= 0\ \text{et}&\nexteqnno[RV]\cr
4\omega_{\overline{z}} s_m + 2 s_{m,\overline{z}} + i u_{m-1} &= 0.&\nexteqnno[RVI]\cr}
$$
\endproclaim
\proclabel{LaxFormulae}
\proof Nous n'allons d\'eduire que les relations provenants de l'\'equation
$$
\Phi_z = [\Phi,\alpha_z],\eqnum{\nexteqnno[KFEI]}
$$
puisque les relations qui proviennent de l'\'equation
$$
\Phi_{\overline{z}} = [\Phi,\alpha_{\overline{z}}]
$$
sont d\'eduites de mani\`ere identique. Notons d'abord
$$
\Phi_m := \pmatrix u_m& e^\omega t_m\cr e^\omega s_m&-u_m\cr\endpmatrix.
$$
En se servant des relations de commutation
$$\eqalign{
[\sigma_-,\sigma_+] &= -\sigma_0,\cr
[\sigma_0,\sigma_-] &= -2\sigma_-\ \text{et}\cr
[\sigma_0,\sigma_+] &= 2\sigma_+,\cr}\eqnum{\nexteqnno[Commutators]}
$$
nous obtenons, pour tout $m$,
$$\eqalign{
[\Phi_m,\alpha_z]
&=\bigg(\frac{i\gamma}{4}t_m - \frac{i}{4\lambda} e^{2\omega}s_m\bigg)\sigma_0\cr
&\qquad+\bigg(-\frac{i\gamma}{2} e^{-\omega} u_m + e^\omega\omega_z s_m\bigg)\sigma_-\cr
&\qquad\qquad+\bigg(\frac{i}{2\lambda} e^\omega u_m - e^\omega\omega_z t_m\bigg)\sigma_+.\cr}
$$
Le coefficient de $\lambda^m$ dans l'expression
$$
\Phi_z - [\Phi,\alpha_z]
$$
devient alors
$$\multiline{
\bigg(u_{m,z} - \frac{i\gamma}{4}t_m + \frac{i}{4} e^{2\omega}s_{m+1}\bigg)\sigma_0\cr
\qquad+\bigg(e^\omega s_{m,z} + \frac{i\gamma}{2} e^{-\omega} u_m\bigg)\sigma_-\cr
\qquad\qquad+\bigg(e^\omega t_{m,z} - \frac{i}{2} e^\omega u_{m+1} + 2 e^\omega\omega_z t_m\bigg)\sigma_+,\cr}
$$
et les relations \eqnref{RI}, \eqnref{RIII} et \eqnref{RV} en d\'ecoulent en identifiant ceci avec zero.\qed
\newsubhead{L'espace de champs de Killing}[LEspaceDeChampsDeKilling]
Notons $\Cal{K}(\Sigma)$ l'espace de champs de Killing sur $\Sigma$. Dans \cite{PinkallSterling}, en construisant un champ de Killing explicite, Pinkall \& Sterling montrent que cet espace est non-trivial\footnote*{Remarquons que le formalisme de \cite{PinkallSterling} est l\'eg\`erement diff\'erent du notre, mais se ram\`ene au notre en rempla\c{c}ant leur variable $z$ avec la variable
$$
\zeta=iz/2
$$
de mani\`ere \`a ce que
$$\eqalign{
\partial_\zeta &= -2i\partial_z\ \text{et}\cr
\partial_{\overline{\zeta}} &= 2i\partial_{\overline{z}}.\cr}
$$}. Ce champ de Killing, que nous appelerons {\sl champ de Pinkall-Sterling} se construit au moyen de la formule r\'ecursive suivante. D'abord, posons
$$\eqalign{
u_0 &:= 0\vphantom{\frac{1}{2}}\ \text{et}\cr
\psi_0 &:= -\frac{1}{2},\cr}\eqnum{\nexteqnno[InitialValuesOfRecursion]}
$$
et, apr\`es avoir d\'etermin\'e $u_1,\cdots,u_m$ et $\psi_1,\cdots,\psi_{m-1}$, posons
$$
\psi_m :=
\left\{
\eqalignleft{
\gamma u_k^2 + 2\sum_{n=1}^{k-1}\theta_{n,m-n}&\qquad\text{si}\ m=2k-1,\ \text{et}\cr
\gamma u_k u_{k+1} + \theta_{k,k} + 2\sum_{n=1}^{k-1}\theta_{n,m-n}&\qquad\text{si}\ m=2k\cr}\right.
\eqnum{\nexteqnno[RecursionForPsiI]}
$$
et
$$
u_{m+1} := \frac{1}{\gamma}\big(-4u_{m,zz} + 4i\omega_z\psi_m\big)\eqnum{\nexteqnno[RecursionForU]}
$$
o\`u, pour tout $p$ et pour tout $q$,
$$
\theta_{p,q} := \gamma u_p u_{q+1} + 4 u_{p,z}u_{q,z} + \psi_p\psi_q.\eqnum{\nexteqnno[RecursionForPsiII]}
$$
Les suites $(t_m)$ et $(s_m)$ sont d\'etermin\'ees par
$$\eqalign{
t_m &:= \frac{1}{\gamma}\big(-2iu_{m,z} - \psi_m)\ \text{et}\cr
s_m &:= e^{-2\omega}\big(2iu_{m-1,z} - \psi_{m-1}\big)\vphantom{\frac{1}{2}},\cr}\eqnum{\nexteqnno[RecursionForTAndS]}%
$$
et le champ de Pinkall-Sterling est la s\'erie
$$
\Phi := \sum_{m=0}^\infty\pmatrix u_m& e^\omega t_m\cr e^\omega s_m& -u_m\cr\endpmatrix\lambda^m.\eqnum{\nexteqnno[ExpansionForPhi]}
$$
\par
\`A l'aide du champ de Pinkall-Sterling nous montrons que $\Cal{K}(\Sigma)$ est de dimension $1$ sur $\Cal{L}$. Pour cela, nous allons nous servir du lemme technique suivant.
\proclaim{Lemme \nextprocno}
\noindent Soit
$$
\Phi := \sum_{m=0}^\infty\pmatrix u_m& e^\omega t_m\cr e^\omega s_m& -u_m\cr\endpmatrix\lambda^m
$$
un champ de Killing sur $\Sigma$. Pour tout $m$,
\medskip
\myitem{(1)} si $u_m=0$, alors $t_m$ et $s_m$ sont constantes;
\medskip
\myitem{(2)} si $u_m=t_m=0$, alors $s_{m+1}=u_{m+1}=0$; et
\medskip
\myitem{(3)} si $u_m=s_m=0$, alors $u_{m-1}=t_{m-1}=0$.
\endproclaim
\proclabel{VanishingConditions}
\proof Supposons que $u_m=0$. Par \eqnref{RIV},
$$
t_{m,\overline{z}} = 0.
$$
Ensuite, par \eqnref{RI},
$$
e^\omega s_{m+1} = \gamma e^{-\omega}t_m,
$$
par \eqnref{RV},
$$\eqalign{
\gamma e^{-\omega}u_{m+1} &= 2ie^\omega s_{m+1,z}\cr
&=2i(e^\omega s_{m+1})_z - 2i\omega_z e^\omega s_{m+1}\cr
&=2i\gamma(e^{-\omega}t_m)_z - 2i\gamma\omega_z e^{-\omega}t_m\cr
&=2i\gamma e^{-\omega}t_{m,z} - 4i\gamma\omega_z e^{-\omega}t_m,\cr}
$$
et par \eqnref{RIII},
$$
t_{m,z} = 0.
$$
Il en d\'ecoule que $t_m$ est constante. Nous montrons de la m\^eme mani\`ere que $s_m$ est constante, et $(1)$ en d\'ecoule. Si $u_m=t_m=0$, il s'ensuit par \eqnref{RI} et \eqnref{RIII} que $s_{m+1}=u_{m+1}=0$, et $(2)$ en d\'ecoule. Si $s_m=u_m=0$, il s'ensuit par \eqnref{RII} et \eqnref{RVI} que $u_{m-1}=t_{m-1}=0$, et $(3)$ en d\'ecoule.\qed
\proclaim{Lemme \nextprocno}
\noindent $\Cal{K}(\Sigma)$ est l'espace vectoriel de dimension $1$ sur $\Cal{L}$ engendr\'e par le champ de Pinkall-Sterling.
\endproclaim
\proclabel{Unidimensional}
\remark En particulier, nous voyons que dans le cadre actuel une solution de l'\'equation sinh-Gordon est de type fini si et seulement si son espace de champs de Killing est engendr\'e par un polyn\^ome de Laurent.
\medskip
\proof Soit $\Phi$ un champ de Killing quelconque et soit $\Psi$ le champ de Pinkall-Sterling. En raisonnant par r\'ecurrence, nous allons montrer l'existence d'une s\'erie de Laurent $\alpha$ qui v\'erifie
$$
\Phi = \alpha\Psi.
$$
En effet, soit $k$ le degr\'e de $\Phi$. Comme $\Psi$ est de degr\'e $0$, la s\'erie $\alpha$ doit aussi \^etre de degr\'e $k$. Supposons maintenant que nous ayons d\'ej\`a d\'etermin\'e les coefficients $\alpha_k,\alpha_{k+1},...,\alpha_{k+l-1}$ de mani\`ere \`a ce que la s\'erie
$$
\tilde{\Phi} := \Phi - \alpha_{(l)}\Psi
$$
soit de degr\'e $k+l$, o\`u
$$
\alpha_{(l)} := \sum_{m=k}^{k+l-1}\alpha_m\lambda^m.
$$
Pour tout $m$, notons
$$
\tilde{\Phi}_m := \pmatrix u_m & e^\omega\tau_m \cr e^\omega\sigma_m & -u_m \cr\endpmatrix.
$$
Comme $\tilde{\Phi}$ est aussi solution de l'equation de champ de Killing, il suit du lemme \procref{VanishingConditions} que
$$\eqalign{
s_{k+l} &=0,\cr
u_{k+l} &=0\ \text{et}\cr
\tau_{k+l} &= c,\cr}
$$
o\`u $c$ est une constante, et nous obtenons le r\'esultat en posant $\alpha_{k+l}:=2c$.\qed
\newsubhead{Le d\'eterminant}[LeDeterminant]
Nous \'etudions maintenant comment le champ de Pinkall-Sterling est caracteris\'e entre tous les champs de Killing. Remarquons d'abord que, comme les \'el\'ements de $\Cal{K}(\Sigma)$ sont des matrices $2\times 2$ \`a coefficients dans $\Cal{L}(\Sigma)$, leurs d\'eterminants sont bien d\'efinis dans $\Cal{L}(\Sigma)$.
\proclaim{Lemme \nextprocno}
\noindent Pour tout champ de Killing $\Phi$, $\opDet(\Phi)$ est constant sur $\Sigma$, c'est-\`a-dire~:
$$
\opDet(\Phi)\in\Cal{L}.
$$
\endproclaim
\proof En effet
$$\eqalign{
d\opDet(\Phi) &= \opTr(\opAdj(\Phi)d\Phi)\cr
&= \opTr(\opAdj(\Phi)[\Phi,\alpha])\cr
&= \opTr(\opAdj(\Phi)\Phi\alpha - \opAdj(\Phi)\alpha\Phi)\cr
&= \opTr(\opAdj(\Phi)\Phi\alpha - \Phi\opAdj(\Phi)\alpha)\cr
&= \opTr(\opDet(\Phi)\alpha - \opDet(\Phi)\alpha)\cr
&= 0,\cr}
$$
et le r\'esultat en d\'ecoule.\qed
\proclaim{Lemme \nextprocno}
\noindent Pour tout $\gamma$, le champ de Pinkall-Sterling est, au signe pr\`es, l'unique champ de Killing $\Phi$ qui v\'erifie
$$
\opDet(\Phi) = -\frac{\lambda}{4\gamma}.\eqnum{\nexteqnno[PSGuage]}
$$
\endproclaim
\proclabel{PSGuage}
\proof Soit $\Phi$ le champ de Pinkall-Sterling. Montrons d'abord l'unicit\'e. Soit $\Psi$ un autre champ de Killing qui v\'erifie \eqnref{PSGuage}. Comme $\Cal{K}(\Sigma)$ est engendr\'e par $\Phi$, il existe une s\'erie de Laurent $\alpha\in\Cal{L}$ telle que
$$
\Psi = \alpha\Phi.
$$
En particulier,
$$
-\frac{\lambda}{4\gamma} = \opDet(\Psi) = \opDet(\alpha\Phi) = -\frac{\lambda}{4\gamma}\alpha^2.
$$
Comme $\Cal{L}$ est un corps alg\'ebrique, nous avons alors
$$\triplealign{
&\alpha^2 - 1&=0\cr
\Leftrightarrow&(\alpha-1)(\alpha+1) &= 0\cr
\Leftrightarrow&\alpha &=\pm 1,\cr}
$$
et l'unicit\'e en d\'ecoule.
\par
Montrons maintenant que $\Phi$ v\'erifie \eqnref{PSGuage}. Pour cela, notons
$$\eqalign{
U &:= \sum_{k=0}^\infty u_k\lambda^k,\cr
S &:= \sum_{k=0}^\infty s_k\lambda^k,\cr
T &:= \sum_{k=0}^\infty t_k\lambda^k\ \text{et}\cr
\Psi &:= \sum_{k=0}^\infty \psi_k\lambda^k,\cr}
$$
o\`u $(\psi_m)$ est la suite construite dans \eqnref{RecursionForPsiI}. Par \eqnref{RecursionForTAndS}
$$\eqalign{
T &= \frac{1}{\gamma}\big(-2iU_z - \Psi\big)\ \text{et}\cr
S &= \lambda e^{-2\omega}\big(2iU_z - \Psi\big),\cr}
$$
et nous obtenons alors
$$\eqalign{
Det(\Phi) &= -U^2 - e^{2\omega}ST\vphantom{\frac{1}{2}}\cr
&= -U^2+\frac{\lambda}{\gamma}(2iU_z - \Psi)(2iU_z + \Psi)\cr
&= -U^2-4\frac{\lambda}{\gamma} U_z^2 - \frac{\lambda}{\gamma}\Psi^2.\cr}
$$
Comme les coefficients de l'expression
$$
-U^2 -4\frac{\lambda}{\gamma} U_z^2 - \frac{\lambda}{\gamma}\Psi^2 = -\frac{\lambda}{4\gamma}
$$
sont pr\'ecis\'ement les relations de r\'ecurrence \eqnref{RecursionForPsiI} et \eqnref{RecursionForPsiII}, le r\'esultat en d\'ecoule.\qed
\newsubhead{La matrice de Sklyanin}[SklyaninsKMatrix]
Nous montrons maintenant comment les conditions au bord de Durham se traduisent en des conditions au bord pour la paire de Lax et pour le champ de Pinkall-Sterling. Remarquons d'abord que la composante r\'eel de la paire de Lax est
$$
\alpha_x = -\frac{i}{2}\omega_y\sigma_0 + \bigg(\frac{i}{4\lambda}e^\omega + \frac{i}{4\gamma}e^{-\omega}\bigg)\sigma_+ + \bigg(\frac{i\gamma}{4}e^{-\omega} + \frac{i\lambda}{4}e^\omega\bigg)\sigma_-.\eqnum{\nexteqnno[LaxPairReal]}
$$
Introduisons maintenant la {\sl matrice de Sklyanin}~:
$$
K := \pmatrix 4A\gamma - 4B\lambda & \frac{\lambda}{\gamma} - \frac{\gamma}{\lambda}\cr
\frac{\lambda}{\gamma} - \frac{\gamma}{\lambda}& \frac{4A}{\gamma} - \frac{4B}{\lambda}\cr\endpmatrix.\eqnum{\nexteqnno[KMatrix]}
$$
Rapellons que $\alpha_x$ v\'erifie la {\sl condition de Sklyanin pour les paires de Lax} en un point de $\partial\Sigma$ lorsque
$$
K(\lambda,\gamma)\alpha_x(\lambda,\gamma) = \alpha_x\bigg(\frac{1}{\lambda},\frac{1}{\gamma}\bigg)K(\lambda,\gamma)\eqnum{\nexteqnno[KMatrixRelation]}
$$
en ce point pour tous $\lambda,\gamma\in S^1$.
\proclaim{Lemme \nextprocno}
\noindent La fonction $\omega$ v\'erifie la condition de Durham
$$
\omega_y = Ae^\omega + Be^{-\omega}
$$
en un point de $\partial\Sigma$ si et seulement si la partie r\'eelle $\alpha_x$ de la paire de Lax v\'erifie la condition de Sklyanin pour les paires de Lax en ce point.
\endproclaim
\proclabel{SklyaninForLaxHolds}
\proof Consid\'erons d'abord une matrice g\'en\'erale de la forme
$$
K := a\opId + b\sigma_0 + c(\sigma_+ + \sigma_-),\eqnum{\nexteqnno[GeneralK]}
$$
\pagebreak
o\`u les coefficients $a$, $b$ et $c$ ne d\'ependent que de $\gamma$ et de $\lambda$. La relation \eqnref{KMatrixRelation} nous donne
$$
-\frac{i}{2}\omega_y[K,\sigma_0] + \bigg(\frac{i}{4\lambda}e^\omega + \frac{i}{4\gamma}e^{-\omega}\bigg)(K\sigma_+-\sigma_-K)
+\bigg(\frac{i\lambda}{4}e^\omega + \frac{i\gamma}{4}e^{-\omega}\bigg)(K\sigma_--\sigma_+K) = 0.
$$
Pourtant,
$$\eqalign{
[K,\sigma_0] &= -2c(\sigma_+-\sigma_-),\cr
K\sigma_+ - \sigma_-K &= (a+b)(\sigma_+-\sigma_-)\ \text{et}\cr
K\sigma_- - \sigma_+K &= -(a-b)(\sigma_+-\sigma_-),\cr}
$$
et \eqnref{KMatrixRelation} devient
$$
c\omega_y + \bigg(\frac{(a+b)}{4\lambda} - \frac{(a-b)\lambda}{4}\bigg)e^\omega + \bigg(\frac{(a+b)}{4\gamma} - \frac{(a-b)\gamma}{4}\bigg)e^{-\omega} = 0.
$$
En posant
$$\eqalign{
a+b &:= 4A\gamma - 4B\lambda,\vphantom{\frac{1}{2}}\cr
a-b &:= \frac{4A}{\gamma} - \frac{4B}{\lambda}\ \text{et}\cr
c &:= \frac{\lambda}{\gamma} - \frac{\gamma}{\lambda},\cr}
$$
nous voyons alors que la relation \eqnref{KMatrixRelation} est v\'erifi\'ee en un point de $\partial\Sigma$ si et seulement si
$$
\omega_y = Ae^\omega + Be^{-\omega}
$$
en ce point, ce qui est pr\'ecis\'ement la condition de Durham. Enfin, en substituant pour $a$, $b$ et $c$ dans \eqnref{GeneralK}, nous obtenons \eqnref{KMatrix}, et le r\'esultat en d\'ecoule.\qed
\medskip
Soit $\partial\Sigma_0$ une des deux composantes connexes de $\partial\Sigma$. Appelons {\sl champ de Killing} sur $\partial\Sigma_0$ toute s\'erie de Laurent $\Phi$ dans $\Cal{L}(\partial\Sigma_0,\opsl(2,\Bbb{C}))$ qui v\'erifie l'\'equation de champ de Killing
$$
\Phi_x = [\Phi,\alpha_x].\eqnum{\nexteqnno[KillingFieldII]}
$$
Notons $\Cal{K}(\partial\Sigma_0)$ l'espace de champs de Killing sur $\partial\Sigma_0$. En particulier, tout champ de Killing sur $\Sigma$ se restreint en un champ de Killing sur $\partial\Sigma_0$.
\proclaim{Lemme \nextprocno}
\noindent Soit
$$
\Phi := \sum_{m=0}^\infty\pmatrix u_m&e^\omega t_m\cr e^\omega s_m&-u_m\endpmatrix\lambda^m
$$
un champ de Killing sur $\partial\Sigma_0$. Supposons que
$$
t_{k-2} = s_{k-1} = u_{k-1} = t_{k-1} = 0,
$$
alors $s_k=u_k=0$ et $t_k$ est constante.
\endproclaim
\proof Par \eqnref{LaxPairReal} et \eqnref{KillingFieldII}, pour tout $m$,
$$\eqalignno{
u_{m,x} &= \frac{i\gamma}{4}t_m + \frac{i}{4}e^{2\omega}t_{m-1} - \frac{i}{4}e^{2\omega}s_{m+1} - \frac{i}{4\gamma}s_m,&\nexteqnno[XRI]\cr
e^\omega s_{m,x} &=-2e^\omega\omega_{\overline{z}}s_m - \frac{i\gamma}{2}e^{-\omega}u_m - \frac{i}{2}e^\omega u_{m-1},\ \text{et}&\nexteqnno[XRII]\cr
e^\omega t_{m,x} &=-2e^\omega\omega_z t_m + \frac{i}{2}e^\omega u_{m+1} + \frac{i}{2\gamma}e^{-\omega}u_m.&\nexteqnno[XRIII]\cr}
$$
Par \eqnref{XRI} and \eqnref{XRIII},
$$
s_k = u_k = 0.
$$
Ces trois relations nous donnent alors
$$\eqalign{
e^{2\omega}s_{k+1} &= \gamma t_k,\vphantom{\frac{1}{1}}\cr
u_{k+1} &= \frac{1}{\gamma}\big(2ie^{2\omega}s_{k+1,x}+4ie^{2\omega}\omega_{\overline{z}}s_{k+1}\big),\ \text{et}\cr
e^\omega t_{k,x} &= -2e^\omega\omega_z t_k + \frac{i}{2}e^\omega u_{k+1}.\vphantom{\frac{1}{2}}\cr}
$$
Il s'ensuit que
$$\eqalign{
u_{k+1} &= 2ie^{2\omega}(e^{-2\omega}t_k)_{,x} + 4i\omega_{\overline{z}}t_k\cr
&=-4i\omega_xt_k + 2it_{k,x} + 4i\omega_{\overline{z}}t_k,\cr}
$$
et donc
$$\triplealign{
&e^\omega t_{k,x} &= -2e^\omega\omega_z t_k + 2e^\omega\omega_x t_k - e^\omega t_{k,x} - 2e^\omega\omega_{\overline{z}}t_k\cr
\Rightarrow & t_{k,x} &= 0,\cr}
$$
et le r\'esultat en d\'ecoule.\qed
\medskip
De la m\^eme mani\`ere que pr\'ec\'edemment, nous obtenons alors le r\'esultat suivant, qui est le pendant dans le cadre actuel des lemmes \procref{Unidimensional} et \procref{PSGuage}~:
\proclaim{Lemme \nextprocno}
\myitem{(1)} $\Cal{K}(\partial\Sigma_0)$ est un espace vectoriel de dimension $1$ sur $\Cal{L}$ engendr\'e par le champ de Pinkall-Sterling, et
\medskip
\myitem{(2)} la restriction du champ de Pinkall-Sterling \`a $\partial\Sigma_0$ est, au signe pr\`es, l'unique champ de Killing $\Phi$ sur $\partial\Sigma_0$ qui v\'erifie
$$
\opDet(\Phi) = -\frac{\lambda}{4\gamma}.
$$
\medskip
\endproclaim
\proclabel{KillingFieldsAlongBoundary}
Pour $\Phi$ un champ de Killing sur $\partial\Sigma_0$, notons
$$
\tilde{\Phi}(\lambda,\gamma) := K(\lambda,\gamma)^{-1}\overline{\Phi\bigg(\frac{1}{\lambda},\frac{1}{\gamma}\bigg)}^tK(\lambda,\gamma).\eqnum{\nexteqnno[PhiTilde]}
$$
Remarquons que, comme les variables $\lambda$ et $\gamma$ sont unitaires, $\tilde{\Phi}$ est aussi une s\'erie de Laurent sur $\partial\Sigma_0$.
\proclaim{Lemme \nextprocno}
\noindent Si la partie r\'eelle $\alpha_x$ de la paire de Lax v\'erifie la condition de Sklyanin pour les paires de Lax le long de $\partial\Sigma_0$, alors $\tilde{\Phi}$ v\'erifie l'\'equation de champ de Killing \eqnref{KillingFieldII} le long de $\partial\Sigma_0$.
\endproclaim
\proclabel{ConjugationByKPreservesKilling}
\proof Comme $\lambda$ et $\gamma$ sont unitaires, nous avons
$$\eqalign{
\overline{\alpha_x(\lambda,\gamma)}^t &= -\alpha_x(\lambda,\gamma),\vphantom{\bigg|}\cr
\overline{K(\lambda,\gamma)} &= D(\lambda,\gamma)K(\lambda,\gamma)^{-1}\ \text{et}\vphantom{\bigg|}\cr
K\bigg(\frac{1}{\lambda},\frac{1}{\gamma}\bigg) &= D(\lambda,\gamma)K(\lambda,\gamma)^{-1},\cr}
$$
o\`u
$$
D(\lambda,\gamma) := \opDet(K(\lambda,\gamma)).
$$
En plus,
$$
D(\lambda,\gamma) = D\bigg(\frac{1}{\lambda},\frac{1}{\gamma}\bigg) = \overline{D(\lambda,\gamma)}.
$$
En appliquant alors la relation de champ de Killing \eqnref{KillingFieldII}, nous obtenons
$$\eqalign{
\tilde{\Phi}(\lambda,\gamma)_x &= K(\lambda,\gamma)^{-1}\overline{\Phi\bigg(\frac{1}{\lambda},\frac{1}{\gamma}\bigg)_x}^tK(\lambda,\gamma)\cr
&= K(\lambda,\gamma)^{-1}\overline{\bigg[\Phi\bigg(\frac{1}{\lambda},\frac{1}{\gamma}\bigg),\alpha_x\bigg(\frac{1}{\lambda},\frac{1}{\gamma}\bigg)\bigg]}^tK(\lambda,\gamma)\cr
&= K(\lambda,\gamma)^{-1}\bigg[\overline{\alpha_x\bigg(\frac{1}{\lambda},\frac{1}{\gamma}\bigg)}^t,\overline{\Phi\bigg(\frac{1}{\lambda},\frac{1}{\gamma}\bigg)}^t\bigg]K(\lambda,\gamma)\cr
&= -K(\lambda,\gamma)^{-1}\bigg[\alpha_x\bigg(\frac{1}{\lambda},\frac{1}{\gamma}\bigg),\overline{\Phi\bigg(\frac{1}{\lambda},\frac{1}{\gamma}\bigg)}^t\bigg]K(\lambda,\gamma)\cr
&= K(\lambda,\gamma)^{-1}\bigg[\overline{\Phi\bigg(\frac{1}{\lambda},\frac{1}{\gamma}\bigg)}^t,\alpha_x\bigg(\frac{1}{\lambda},\frac{1}{\gamma}\bigg)\bigg]K(\lambda,\gamma)\cr
&= \bigg[K(\lambda,\gamma)^{-1}\overline{\Phi\bigg(\frac{1}{\lambda},\frac{1}{\gamma}\bigg)}^tK(\lambda,\gamma),K(\lambda,\gamma)^{-1}\alpha_x\bigg(\frac{1}{\lambda},\frac{1}{\gamma}\bigg)K(\lambda,\gamma)\bigg]\cr
&= [\tilde{\Phi}(\lambda,\gamma),\alpha_x(\lambda,\gamma)],\vphantom{\bigg[}\cr}
$$
et le r\'esultat en d\'ecoule.\qed
\medskip
\noindent Rappelons que $\Phi$ v\'erifie la {\sl condition de Sklyanin pour les champs} en un point de $\partial\Sigma_0$ lorsque
$$
\Phi(\lambda,\gamma) = \tilde{\Phi}(\lambda,\gamma)
$$
en ce point pour tous $\lambda,\gamma\in S^1$.
\proclaim{Lemme \nextprocno}
\noindent Si la partie r\'eelle $\alpha_x$ de la paire de Lax v\'erifie la condition de Sklyanin pour les paires de Lax le long de $\partial\Sigma_0$, alors le champ de Pinkall-Sterling $\Phi$ v\'erifie la condition de Sklyanin pour les champs le long de $\partial\Sigma_0$, c'est-\`a-dire, pour tout $\lambda,\gamma\in S^1$,
$$
\Phi(\lambda,\gamma) = \tilde{\Phi}(\lambda,\gamma)\eqnum{\nexteqnno[KReality]}
$$
le long de $\partial\Sigma$.
\endproclaim
\proclabel{PinkalSterlingFieldIsReal}
\proof Par le lemme \procref{ConjugationByKPreservesKilling}, $\tilde{\Phi}$ est un champ de Killing sur $\partial\Sigma$. Or, pour tout $\lambda$ et pour tout $\gamma$,
$$\eqalign{
\opDet(\tilde{\Phi}(\lambda,\gamma)) &= \opDet\bigg(K(\lambda,\gamma)^{-1}\overline{\Phi\bigg(\frac{1}{\lambda},\frac{1}{\gamma}\bigg)}^tK(\lambda,\gamma)\bigg)\cr
&=\opDet\bigg(\overline{\Phi\bigg(\frac{1}{\lambda},\frac{1}{\gamma}\bigg)}\bigg)\cr
&=\overline{\opDet\bigg(\Phi\bigg(\frac{1}{\lambda},\frac{1}{\gamma}\bigg)\bigg)}\cr
&=-\frac{\lambda}{4\gamma}.\cr}
$$
Il en d\'ecoule par le lemme \procref{KillingFieldsAlongBoundary} que
$$
\Phi = \pm\tilde{\Phi},
$$
et le r\'esultat en d\'ecoule en d\'eterminant explicitement le premier terme non-nul de chacune de ces deux s\'eries.\qed
\newsubhead{Les conditions au bord de Robin}[LaConditionDeBordDeRobin]
Nous allons transformer maintenant la condition de Sklyanin pour des champs en des \'equations qui vont nous permettre de montrer que les solutions sont de type fini. Soit $\Phi$ alors le champ de Pinkall-Sterling. Afin de mieux percevoir les symm\'etries du probl\`eme, il convient maintenant de consid\'erer ce champ-ci comme une s\'erie de Laurent en $\lambda$ et $\gamma$. Notons alors
$$
\Phi(\lambda,\gamma) := \sum_{m,n}\pmatrix u_{m,n}& e^\omega t_{m,n}\cr e^\omega s_{m,n} & -u_{m,n}\cr\endpmatrix\lambda^m\gamma^n.\eqnum{\nexteqnno[NewExpansionForPhi]}
$$
En fait, par \eqnref{GaugeTransform},
$$
\Phi(\lambda,\gamma) = \sum_{m=0}^\infty\pmatrix u_m&\frac{1}{\gamma}e^\omega t_m\cr \gamma e^\omega s_m& -u_m\cr\endpmatrix\lambda^m\gamma^{-m},\eqnum{\nexteqnno[PSFieldGeneralGamma]}
$$
o\`u
$$
\sum_{m=0}^\infty\pmatrix u_m& e^\omega t_m\cr e^\omega s_m& -u_m\cr\endpmatrix\lambda^m := \Phi(\lambda,1)
$$
est le champ Pinkall-Sterling comme param\`etre de torsion $\gamma=1$.
\par
Dans ce cadre, le lemme \procref{LaxFormulae} devient~:
\proclaim{Lemme \nextprocno}
\noindent Les suites $(u_{m,n})$, $(t_{m,n})$ et $(s_{m,n})$ v\'erifient, pour tout $m$ et pour tout $n$,
$$\eqalignno{
4u_{m,n,z} + i e^{2\omega} s_{m+1,n} - i t_{m,n-1} &=0, &\nexteqnno[NRI]\cr
4u_{m,n,\overline{z}} + is_{m,n+1} - ie^{2\omega} t_{m-1,n} &=0, &\nexteqnno[NRII]\cr
4\omega_z t_{m,n} + 2 t_{m,n,z} - i u_{m+1,n} &=0, &\nexteqnno[NRIII]\cr
2e^\omega t_{m,n,\overline{z}} - i e^{-\omega}u_{m,n+1} &=0, &\nexteqnno[NRIV]\cr
2e^\omega s_{m,n,z} + i e^{-\omega} u_{m,n-1} &=0\ \text{et} &\nexteqnno[NRV]\cr
4\omega_{\overline{z}}s_{m,n} + 2s_{m,n,\overline{z}} + iu_{m-1,n} &=0.&\nexteqnno[NRVI]\cr}
$$
\endproclaim
\noindent En ce qui concerne la condition de Sklyanin pour les champs, nous avons
\proclaim{Lemme \nextprocno}
\noindent Le long de $\partial\Sigma$, les suites $(u_{m,n})$, $(t_{m,n})$ et $(s_{m,n})$ v\'erifient, pour tout $m$ et pour tout $n$,
$$\eqalignno{
\opIm\big(e^\omega s_{m-1,n+1} - e^\omega s_{m+1,n-1} + 4Au_{m,n-1} - 4Bu_{m-1,n}\big) &= 0,&\nexteqnno[SI]\cr
\opIm\big(e^\omega t_{m-1,n+1} - e^\omega t_{m+1,n-1} - 4Au_{m,n+1} + 4Bu_{m+1,n}\big) &= 0,&\nexteqnno[SII]\cr
\opRe\big(2Ae^\omega t_{m,n-1} - 2Be^\omega t_{m-1,n} - 2Ae^\omega s_{m,n+1} + 2Be^\omega s_{m+1,n}\big) & &\cr
\qquad\qquad\qquad=\opRe\big(u_{m-1,n+1} - u_{m+1,n-1}\big)\ \text{et}& &\nexteqnno[SIII]\cr
\opIm\big(At_{m,n-1} - Bt_{m-1,n} + As_{m,n+1} - Bs_{m+1,n}\big) &= 0.&\nexteqnno[SIV]\cr}
$$
\endproclaim
\proof En effet, notons
$$
\hat{\Phi}(\lambda,\gamma) := \overline{\Phi\bigg(\frac{1}{\lambda},\frac{1}{\gamma}\bigg)}^t =
\sum_{m,n}\pmatrix \overline{u}_{m,n}&e^\omega\overline{s}_{m,n}\cr e^\omega\overline{t}_{m,n}&-\overline{u}_{m,n}\cr\endpmatrix\lambda^m\gamma^n.
$$
Comme $\Phi$ v\'erifie la condition de Sklyanin pour les champs, tous les coefficients de la s\'erie
$$
K(\lambda,\gamma)\Phi(\lambda,\gamma) - \hat{\Phi}(\lambda,\gamma)K(\lambda,\gamma)
$$
s'annulent, et le r\'esultat en d\'ecoule.\qed
\proclaim{Lemme \nextprocno}
\noindent La paire de syst\`emes de relations \eqnref{SI} et \eqnref{SII} est \'equivalente \`a la paire de syst\`emes de relations suivante~:
$$\eqalignno{
\opIm\big(e^\omega t_{m-1,n} - 4Au_{m,n} + e^\omega s_{m+1,n}\big) &= 0\ \text{et}&\nexteqnno[NSI]\cr
\opIm\big(e^\omega t_{m,n-1} - 4Bu_{m,n} + e^\omega s_{m,n+1}\big) &= 0.&\nexteqnno[NSII]\cr}
$$
\endproclaim
\proof En effet, supposons que les relations \eqnref{SI} et \eqnref{SII} soient satisfaites pour tout $m$ et pour tout $n$. Alors, en appliquant r\'ecursivement \eqnref{SI}, nous obtenons la somme t\'el\'escopique finie suivante~:
$$
\opIm\big(e^\omega s_{m,n}\big) = \opIm\big(4A u_{m-1,n} - 4B u_{m-2,n+1} + 4Au_{m-3,n+2} - \cdots\big).
$$
De la m\^eme mani\`ere, de la relation \eqnref{SII} nous obtenons
$$
\opIm\big(e^\omega t_{m,n}\big) = \opIm\big(4B u_{m,n+1} - 4A u_{m-1,n+2} + 4B u_{m-2,n+3} - \cdots\big).
$$
Si nous notons respectivement ces relations $\alpha(m,n)$ et $\beta(m,n)$, alors $\alpha(m+1,n)$ et $\beta(m-1,n)$ nous donnent \eqnref{NSI} alors que $\alpha(m,n+1)$ et $\beta(m,n-1)$ nous donnent \eqnref{NSII}. Comme la r\'eciproque est triviale, le r\'esultat en d\'ecoule.\qed
\proclaim{Lemme \nextprocno}
\noindent Pour tout $m$ et pour tout $n$, la partie imaginaire de la fonction $u_{m,n}$ v\'erifie les conditions au bord de Robin suivantes~:
$$
\opIm(u_{m,n})_y = Ae^\omega\opIm(u_{m,n}) - Be^{-\omega}\opIm(u_{m,n}).\eqnum{\nexteqnno[RobinBoundaryCondition]}
$$
\endproclaim
\proclabel{BoundaryConditionsForImaginaryPart}
\proof En effet, par \eqnref{NRI} et \eqnref{NRII},
$$\eqalign{
e^\omega s_{m+1,n} - e^{-\omega}t_{m,n-1} &= 4ie^{-\omega}u_{m,n,z}\ \text{et}\cr
e^\omega t_{m-1,n} - e^{-\omega}s_{m,n+1} &= -4ie^{-\omega}u_{m,n,\overline{z}}.\cr}
$$
En sommant \eqnref{NSI} et \eqnref{NSII}, nous obtenons alors
$$\eqalign{
\opIm\big(4A u_{m,n} - 4B e^{-2\omega} u_{m,n}\big)
&=\opIm\big(e^\omega t_{m-1,n} + e^\omega s_{m+1,n} - e^{-\omega} t_{m,n-1} - e^{-\omega} s_{m,n+1}\big)\cr
&=\opIm\big(4ie^{-\omega}u_{m,n,z} - 4ie^{-\omega}u_{m,n,\overline{z}}\big)\cr
&=4e^{-\omega}\opRe\big((u_{m,n} - \overline{u}_{m,n})_z\big)\cr
&=4e^{-\omega}\opRe\big(2i\opIm(u_{m,n})_z\big)\cr
&=4e^{-\omega}\opIm(u_{m,n})_y,\cr}
$$
et le r\'esultat en d\'ecoule.\qed
\newsubhead{La solution est de type fini}[LaSolutionEstDeTypeFini]
Nous montrons maintenant que la solution est de type fini. Soit $\Phi$ comme dans la section pr\'ec\'edente. Rappelons d'abord quelques lemmes \'el\'ementaires.
\proclaim{Lemme \nextprocno}
\noindent Soit
$$
\Psi := \sum_{m=0}^\infty\pmatrix u_m&e^\omega t_m\cr e^\omega s_m&-u_m\cr\endpmatrix\lambda^m
$$
un champ de Killing. Si $u_k=0$, alors, il existe une s\'erie de Laurent $\alpha\in\Cal{L}$ de degr\'e $k$ telle que
$$
\Psi - \alpha\Phi = \sum_{m=0}^{k-1}\pmatrix u_m&e^\omega t_m\cr e^\omega s_m&-u_m\cr\endpmatrix\lambda^m + \pmatrix 0&0\cr e^\omega s_k&0\cr\endpmatrix\lambda^k.
$$
En particulier, $(\Psi-\alpha\Phi)$ est un champ de Killing polyn\^omial.
\endproclaim
\proclabel{PreFiniteType}
\proof Nous construisons $\alpha$ par r\'ecurrence. Supposons alors que nous ayons d\'ej\`a d\'etermin\'e les coefficients $\alpha_k,\cdots,\alpha_{k+l-1}$ de mani\`ere \`a ce que si
$$
\tilde{\Psi} := \Psi - \alpha_{(l)}\Phi := \sum_{m=0}^\infty\pmatrix\tilde{u}_{m}&e^\omega\tilde{t}_{m}\cr e^\omega\tilde{s}_{m}&-\tilde{u}_{m}\endpmatrix\lambda^m,
$$
o\`u
$$
\alpha_{(l)} := \sum_{m=k}^{k+l-1}\alpha_m\lambda^m,
$$
alors,
$$\eqalign{
\tilde{u}_{m} &=0\ \forall k\leq m\leq k+l,\cr
\tilde{t}_{m} &=0\ \forall k\leq m\leq k+l-1,\ \text{et}\cr
\tilde{s}_{m} &=0\ \forall k+1\leq m\leq k+l.\cr}
$$
Comme $\tilde{\Psi}$ est aussi un champ de Killing, par le lemme \procref{VanishingConditions},
$$
\tilde{t}_{k+l} = c
$$
est constante. Par le lemme \procref{VanishingConditions} de nouveau, nous obtenons le r\'esultat en posant $\alpha_{k+l}:=-2c$.\qed
\medskip
Rappelons de l'introduction qu'une solution $\omega$ de l'\'equation sinh-Gordon est dite de type fini si et seulement si elle admet un champ de Killing polyn\^omial.
\proclaim{Lemme \nextprocno}
\noindent La solution $\omega$ est de type fini si et seulement s'il existe un sous-espace vectoriel de dimension finie $E\subseteq C^\infty(\Sigma,\Bbb{C})$ tel que, pour tout $m$ et pour tout $n$,
$$
u_{m,n}\in E.
$$
\endproclaim
\proclabel{FiniteTypeCriteria}
\proof Il suffit de montrer que cette condition est suffisante. Supposons de nouveau que $\gamma=1$. En particulier, par \eqnref{PSFieldGeneralGamma}, pour tout $m$, nous avons
$$
u_m = u_{m,m} \in E.
$$
Si $d:=\opDim(E)$, alors il existe un polyn\^ome de Laurent $\alpha\in\Cal{P}$ de bidegr\'e $(0,d-1)$ tel que si
$$
\alpha\Phi := \sum_{m=0}^\infty\pmatrix\tilde{u}_m& e^\omega\tilde{t}_m\cr e^\omega\tilde{s}_m&-\tilde{u}_m\endpmatrix\lambda^m,
$$
alors
$$
\tilde{u}_d = 0,
$$
et le r\'esultat en d\'ecoule par le lemme \procref{PreFiniteType}.\qed
\proclaim{Lemme \nextprocno}
\noindent Pour tout $m$ et pour tout $n$, la fonction $u_{m,n}$ v\'erifie l'\'equation sinh-Gordon lin\'earis\'ee suivante~:
$$
\Delta u_{m,n} + \opCosh(2\omega)u_{m,n} = 0.\eqnum{\nexteqnno[LinearisedSinhGordon]}
$$
\endproclaim
\proof En effet, en d\'erivant \eqnref{NRI}, nous obtenons
$$
4u_{m,n,z\overline{z}} + i(e^{2\omega}s_{m+1,n})_{\overline{z}} - it_{m,n-1,\overline{z}} = 0.
$$
En appliquant \eqnref{NRIV} et \eqnref{NRVI}, ceci devient
$$
4u_{m,n,z\overline{z}} + \frac{1}{2}e^{2\omega}u_{m,n} + \frac{1}{2}e^{-2\omega} u_{m,n} = 0,
$$
et le r\'esultat en d\'ecoule.\qed
\proclaim{Lemme \nextprocno}
\noindent Si $\phi:\Sigma\rightarrow\Bbb{C}$ est une fonction holomorphe qui v\'erifie
$$
\opIm(\phi)|_{\partial\Sigma} = 0,
$$
alors $\phi$ est constante.
\endproclaim
\proclabel{Liouville}
\proof En effet, par le principe de r\'eflexion de Cauchy, $\phi$ s'\'etend en une fonction holomorphe et born\'ee sur le cylindre parabolique $S^1\times\Bbb{R}$ et le r\'esultat en d\'ecoule par le principe de Liouville.\qed
\proclaim{Th\'eor\`eme \nextprocno}
\noindent Soit $\omega:\Sigma\rightarrow\Bbb{R}$ solution de l'\'equation sinh-Gordon avec conditions au bord de Durham. Alors $\omega$ est de type fini.
\endproclaim
\proof En effet, soit $\Phi$ le champ de Pinkall-Sterling de $\omega$ et soient $(u_{m,n})$, $(t_{m,n})$ et $(s_{m,n})$ comme dans \eqnref{NewExpansionForPhi}. Par les lemmes \procref{SklyaninForLaxHolds} et \procref{PinkalSterlingFieldIsReal}, $\Phi$ v\'erifie la condition de Sklyanin pour les champs le long de $\partial\Sigma$. Il en d\'ecoule par le lemme \procref{BoundaryConditionsForImaginaryPart} que, pour tout $m$ et pour tout $n$,
$$
\opIm(u_{m,n})_{,y} = Ae^\omega\opIm(u_{m,n}) - Be^{-\omega}\opIm(u_{m,n}),
$$
le long de $\partial\Sigma$. Comme $\opIm(u_{m,n})$ v\'erifie aussi la lin\'earis\'ee de l'equation sinh-Gordon, il s'ensuit par la th\'eorie classique d'op\'erateurs elliptiques sur des vari\'et\'es compactes \`a bord (c.f. \cite{GilbTrud}) qu'il existe un sous-espace vectoriel de dimension finie $E_1\subseteq C^\infty(\Sigma,\Bbb{R})$ tel que, pour tout $m$ et pour tout $n$,
$$
\opIm(u_{m,n}) \in E_1.
$$
\par
Par \eqnref{NRIII} et \eqnref{NRVI}, pour tout $m$ et pour tout $n$,
$$
\big(e^{2\omega}\overline{s}_{m+1,n} - e^{2\omega}t_{m-1,n}\big)_z
=-\frac{i}{2}e^{2\omega}\big(u_{m,n} - \overline{u}_{m,n}\big)=e^{2\omega}\opIm(u_{m,n})\in e^{2\omega}E_1,
$$
et, par \eqnref{NSI}, le long de $\partial\Sigma$,
$$
\opIm\big(e^{2\omega}\overline{s}_{m+1,n} - e^{2\omega}t_{m-1,n}\big)=-4Ae^\omega\opIm(u_{m,n})\in e^\omega E_1|_{\partial\Sigma}.
$$
Il en d\'ecoule par le lemme \procref{Liouville} qu'il existe un sous-espace vectoriel de dimension finie $E_2\subseteq C^\infty(\Sigma,\Bbb{C})$ tel que pour tout $m$ et pour tout $n$,
$$
e^{2\omega}\overline{s}_{m+1,n} - e^{2\omega}t_{m-1,n} \in e^\omega E_2.
$$
\par
Par \eqnref{NRIV} et \eqnref{NRV}, pour tout $m$ et pour tout $n$,
$$
\big(s_{m,n+1} - \overline{t}_{m,n-1}\big)_z=-\frac{i}{2}e^{-2\omega}\big(u_{m,n} - \overline{u}_{m,n}\big)
=e^{-2\omega}\opIm(u_{m,n})\in e^{-2\omega}E_1.
$$
et, par \eqnref{NSII}, le long de $\partial\Sigma$,
$$
\opIm\big(s_{m,n+1} - \overline{t}_{m,n-1}\big)= 4Be^{-\omega}\opIm(u_{m,n})\in e^{-\omega}E_1|_{\partial\Sigma}.
$$
Il en d\'ecoule de nouveau par le lemme \procref{Liouville} qu'il existe un sous-espace vectoriel de dimension finie $E_3\subseteq C^\infty(\Sigma,\Bbb{C})$ tel que pour tout $m$ et pour tout $n$,
$$
s_{m,n+1} - \overline{t}_{m,n-1} \in e^{-\omega}E_3.
$$
\par
Enfin, par \eqnref{SIII}, le long de $\partial\Sigma$, pour tout $m$ et pour tout $n$,
$$\eqalign{
\opRe\big(u_{m+1,n-1} - u_{m-1,n+1}\big)
&=2Be^{-\omega}\opRe\big(e^{2\omega}\overline{s}_{m+1,n}-e^{2\omega}t_{m-1,n}\big)\cr
&\qquad\quad-2Ae^\omega\opRe\big(s_{m,n+1} - \overline{t}_{m,n-1}\big)\cr
&\in\opRe(E_2+E_3).\cr}
$$
Il en d\'ecoule par r\'ecurrence que, le long de $\partial\Sigma$, pour tout $m$ et pour tout $n$,
$$
\opRe(u_{m,n}) \in \opRe(E_2+E_3).
$$
Enfin, comme $\opRe(u_{m,n})$ v\'erifie aussi la lin\'earis\'ee de l'equation sinh-Gordon, il s'ensuit de nouveau par la th\'eorie classique des op\'erateurs elliptiques sur les vari\'et\'es compactes \`a bord qu'il existe un quatri\`eme sous-espace vectoriel de dimension finie $E_4\subseteq C^\infty(\Sigma,\Bbb{R})$ tel que, pour tout $m$ et pour tout $n$,
$$
\opIm(u_{m,n}) \in E_4.
$$
Le r\'esultat en d\'ecoule par le lemme \procref{FiniteTypeCriteria}.\qed
\newsubhead{Des champs de Killing polynomiaux}[DesChampsDeKillingPolynomes]
Enfin, afin de construire des champs de Killing polynomiaux qui v\'erifient la condition de Sklyanin pour des champs, nous nous restreignons de nouveau au cas o\`u $\gamma=1$.
\proclaim{Lemme \nextprocno}
\noindent Soit
$$
\Psi := \sum_{m=0}^\infty\pmatrix u_m&e^\omega t_m\cr e^\omega s_m&- u_m\cr\endpmatrix\lambda^m
$$
un champ de Killing qui v\'erifie la condition de Sklyanin pour des champs. Si $u_k=0$ et si $t_k\notin\Bbb{R}$ alors il existe un champ de Killing polynomial $\Psi'$ de bidegr\'e $(0,4)$ qui v\'erifie aussi la condition de Sklyanin pour des champs.
\endproclaim
\proclabel{ExceptionalKillingField}
\proof Soit $\Phi$ le champ de Pinkall-Sterling. Par le lemme \procref{PreFiniteType}, il existe des s\'eries de Laurent \`a coefficients r\'eels $f,g\in\Cal{L}$ telles que
$$
\Psi - (f+ig)\Phi = P,
$$
o\`u $P$ est un champ de Killing polynomial. Notons alors
$$
\Psi_1 := g\Phi.
$$
Comme $\Phi$ v\'erifie la condition de Sklyanin pour les champs, et comme $g$ est \`a coefficients r\'eels, $\Psi_1$ v\'erifie aussi la condition de Sklyanin pour les champs, c'est-\`a-dire, pour tout $\lambda\in S^1$,
$$
K(\lambda,1)\Psi_1(\lambda,1) - \overline{\Psi_1\bigg(\frac{1}{\lambda},1\bigg)}^tK(\lambda,1)=0
$$
le long de $\partial\Sigma$. Ensuite, comme $\Phi$ et $\Psi$ v\'erifient tous les deux la condition de Sklyanin pour les champs, et comme $f$ est \`a coefficients r\'eels, nous avons aussi, pour tout $\lambda\in S^1$,
$$
iK(\lambda,1)\Psi_1(\lambda,1)+i\overline{\Psi_1\bigg(\frac{1}{\lambda},1\bigg)}^tK(\lambda,1)=Q(\lambda,1)
$$
le long de $\partial\Sigma$, o\`u
$$
Q(\lambda,1):=K(\lambda,1)P(\lambda,1) - \overline{P\bigg(\frac{1}{\lambda},1\bigg)}^tK(\lambda,1).
$$
Il s'ensuit que
$$\triplealign{
&K(\lambda,1)\Psi_1(\lambda,1) &= -\frac{i}{2}Q(\lambda,1)\cr
\Leftrightarrow &D(\lambda,1)\Psi_1(\lambda,1) &= -\frac{i}{2}K\bigg(\frac{1}{\lambda},1\bigg)Q(\lambda,1),\cr}
$$
o\`u
$$
D(\lambda,1) := \opDet(K(\lambda,1)).
$$
Notons alors
$$
\Psi'(\lambda,1) := \lambda^{2-k}D(\lambda,1)\Psi_1(\lambda,1).
$$
Comme $D(\lambda,1)$ est un polyn\^ome de Laurent \`a coefficients r\'eels, $\Psi'$ est aussi un champ de Killing qui v\'erifie la condition de Sklyanin pour les champs. Enfin, nous v\'erifions que $Q(\lambda,1)$ est un polyn\^ome de bidegr\'e $(k-1,k+1)$, et le r\'esultat en d\'ecoule.\qed
\medskip
{\bf\noindent D\'emonstration du th\'eor\`eme \procref{MainTheorem}~:}\ Pour montrer l'existence d'un champ de Killing polynomial qui v\'erifie la condition de Sklyanin pour les champs, il suffit de refaire la construction des lemmes \procref{PreFiniteType} et \procref{FiniteTypeCriteria} en n'utilisant que des s\'eries de Laurent \`a coefficients r\'eels. La seule obstruction possible \`a cette construction, c'est le cas \'etudi\'e dans le lemme \procref{ExceptionalKillingField}. Comme il existe, de toute fa\c{c}on, un champ de Killing polynomial qui v\'erifie la condition de Sklyanin pour les champs m\^eme dans ce cas-ci, l'existence en d\'ecoule.
\par
R\'eciproquement, supposons qu'il existe un champ de Killing polynomial
$$
\Psi := \sum_{m=0}^k\pmatrix u_m&e^\omega t_m\cr e^\omega s_m&-u_m\cr\endpmatrix\lambda^m
$$
qui v\'erifie la condition de Sklyanin pour les champs. Alors, par le lemme \procref{VanishingConditions},
$$
s_0 = u_0 = 0.
$$
Par \eqnref{SII}, nous pouvons supposer que
$$
t_0 = \frac{1}{2},
$$
et, par \eqnref{SIII},
$$
\opRe\big(u_1 + 2Ae^\omega t_0 + 2Be^\omega s_1\big) = 0.
$$
Or, par \eqnref{InitialValuesOfRecursion}, \eqnref{RecursionForU} et \eqnref{RecursionForTAndS},
$$\eqalign{
u_1 &= -2i\omega_z,\vphantom{\frac{1}{2}}\cr
t_0 &=\frac{1}{2}\ \text{et}\cr
s_0 &=\frac{1}{2}e^{-2\omega},\cr}
$$
et le r\'esultat en d\'ecoule.\qed
\bigskip
\inappendicestrue
\global\headno=0
\goodbreak
\newhead{R\'ef\'erences}[Bibliography]
{\leftskip = 5ex \parindent = -5ex
\leavevmode\hbox to 4ex{\hfil \cite{AdlerEtAl}}\hskip 1ex{Adler V., G\"urel B., G\"urses M., Habublin I., Boundary conditions for integrable equations, {\sl J. Phys. A Math. Gen.}, {\bf 30}, (1997), 3505--3513}
\medskip
\leavevmode\hbox to 4ex{\hfil \cite{BajnokEtAl}}\hskip 1ex{Bajnok Z., Palla L., Tak\'acs G., Boundary sine-Gordon model, arXiv:hep-th/0211132}
\medskip
\leavevmode\hbox to 4ex{\hfil \cite{BobenkoBook}}\hskip 1ex{Belokolos A. D., Bobenko A. I., Enol'skii V. Z., Its A. R., Matveev V. B., {\sl Algebro-geometric approach to nonlinear integrable equations}, Springer-Verlag, (1994)}
\medskip
\leavevmode\hbox to 4ex{\hfil \cite{Bobenko}}\hskip 1ex{Bobenko A. I., All constant mean curvature tori in $\Bbb{R}^3$, $\Bbb{S}^3$, $\Bbb{H}^3$ in terms of theta-functions, {\sl Math. Ann.}, {\bf 290}, (1991), 209--245}
\medskip
\leavevmode\hbox to 4ex{\hfil \cite{BobenkoKuksin}}\hskip 1ex{Bobenko A. I., Kuksin S. B., Small amplitude solutions of the sine-Gordon equation on an interval under Dirichlet or Neumann boundary conditions, {\sl J. Nonlinear Sci.}, {\bf 5}, (1995), 207--232}
\medskip
\leavevmode\hbox to 4ex{\hfil \cite{BurstallFerusPeditPinkall}}\hskip 1ex{Burstall F. E., Ferus D., Pedit F., Pinkall U., Harmonic tori in symmetric spaces and commuting hamiltonian systems on loop algebras, {\sl Ann. of Math.}, {\bf 138}, no. 1, (1993), 173--212}
\medskip
\leavevmode\hbox to 4ex{\hfil \cite{CorriganA}}\hskip 1ex{Corrigan E., Recent developments in affine Toda quantum field theory, in {\sl Particles and Fields}, CRM Series in Mathematical Physics, Springer, New York, NY, (1991)}
\medskip
\leavevmode\hbox to 4ex{\hfil \cite{CorriganEtAlA}}\hskip 1ex{Corrigan E., Dorey P. E., Rietdijk R. H., Sasaki R., Affine Toda field theory on a half line, {\sl Phys. Lett.} {\bf B}, {\bf 333}, (1994)}
\medskip
\leavevmode\hbox to 4ex{\hfil \cite{CorriganEtAlB}}\hskip 1ex{Corrigan E., Dorey P. E., Rietdijk R. H., Aspects of affine Toda field theory on a half line, {\sl Prog. Theor. Phys. Suppl.}, {\bf 118}, (1995), 143--164}
\medskip
\leavevmode\hbox to 4ex{\hfil \cite{Delius}}\hskip 1ex{Delius G. W., Soliton preserving boundary condition in affine Toda field theories, {\sl Phys. Lett.} {\bf B}, {\bf 444}, (1998), 217--223}
\medskip
\leavevmode\hbox to 4ex{\hfil \cite{FadeevTakhtajan}}\hskip 1ex{Fadeev L. D., Takhtajan L., {\sl Hamiltonian methods in the theory of solitons}, Classics in Mathematics, Spring-Verlag, Berling, Heidelberg, (2007)}
\medskip
\leavevmode\hbox to 4ex{\hfil \cite{FokasIts}}\hskip 1ex{Fokas A. S., Its. A. R., An initial value problem for the sine-Gordon equation in laboratory coordinates, {\sl Theor. Math. Phys.}, {\bf 92}, no. 3, 964--978}
\medskip
\leavevmode\hbox to 4ex{\hfil \cite{Fokas}}\hskip 1ex{Fokas A. S., Linearizable initial value problems for the sine-Gordon equation on the half-line, {\sl Nonlinearity}, {\bf 17}, (2004), 1521--1534}
\medskip
\leavevmode\hbox to 4ex{\hfil \cite{FordyWood}}\hskip 1ex{Fordy A. P., Wood J. C., {\sl Harmonic maps and integrable systems}, Aspects of Mathematics, {\bf E23}, Vieweg, Braunschweig, (1994)}
\medskip
\leavevmode\hbox to 4ex{\hfil \cite{GilbTrud}}\hskip 1ex{Gilbarg D., Trudinger N. S., {\sl Elliptic partial differential equations of second order}, Classics in Mathematics, Springer-Verlag, Berlin, (2001)}
\medskip
\leavevmode\hbox to 4ex{\hfil \cite{HauswirthKillianSchmidt}}\hskip 1ex{Hauswirth L., Killian M., Schmidt M., Finite type minimal annuli in $\Bbb{S}^2\times\Bbb{R}$, {\sl Illinois J. Math.}, {\bf 57}, no. 3, (2013), 697--741}
\medskip
\leavevmode\hbox to 4ex{\hfil \cite{Hitchin}}\hskip 1ex{Hitchin N., Harmonic maps from a $2$-torus to the $3$-sphere, {\sl J. Diff. Geom.}, {\bf 31}, no. 3, (1990), 627--710}
\medskip
\leavevmode\hbox to 4ex{\hfil \cite{MacIntyre}}\hskip 1ex{MacIntyre A., Integrable boundary conditions for classical sine-Gordon theory, {\sl J. Phys. A. Math. Gen.}, {\bf 28}, (1995), 1089--1100}
\medskip
\leavevmode\hbox to 4ex{\hfil \cite{PinkallSterling}}\hskip 1ex{Pinkall U., Sterling I, On the Classification of Constant Mean Curvature Tori, {\sl Ann. of Math.}, {\bf 130}, no. 2, (1989), 407--451}
\medskip
\leavevmode\hbox to 4ex{\hfil \cite{SklyaninI}}\hskip 1ex{Sklyanin E. K., Boundary conditions for integrable equations, {\sl Funct. Anal. Appl.}, {\bf 21}, (1987), 164--166}
\medskip
\leavevmode\hbox to 4ex{\hfil \cite{SklyaninII}}\hskip 1ex{Sklyanin E. K., Boundary conditions for integrable quantum systems, {\sl J. Phys. A. Math. Gen.}, {\bf 21}, (1988), 2375--2389}
\par}
%
%
%
%
\enddocument

%% file: preamble.tex
%
%
%
%
\let\myfrac=\frac%
\input eplain %
\let\frac=\myfrac%
\let\myfootnote=\footnote%
\input amstex \input epsf %
\let\footnote=\myfootnote%
%
%
\loadeufm\loadmsam\loadmsbm\message{symbol names}\UseAMSsymbols\message{,}%
\magnification 1200 %
\font\myfontdefault=cmr10%
\newif\ifmakebiblio%
\newif\ifinappendices%
\newif\ifundefinedreferences%
\newif\ifchangedreferences%
\makebibliofalse%
\undefinedreferencesfalse%
\changedreferencesfalse%
%
%
%
%
%
\def\setcatcodes{\catcode`\!=0 \catcode`\\=11}%
{\global\let\noe=\noexpand%
\catcode`\@=11 \catcode`\_=11 \setcatcodes%
!global!def!_@@internal@@makeref#1{%
!global!expandafter!def!csname #1ref!endcsname##1{%
!csname _@#1@##1!endcsname%
!expandafter!ifx!csname _@#1@##1!endcsname!relax%
    !write16{#1 ##1 not defined - run saving references}%
    !undefinedreferencestrue%
!fi}}%
!global!def!_@@internal@@makelabel#1{%
!global!expandafter!def!csname #1label!endcsname##1{%
!edef!temptoken{!csname #1info!endcsname}%
!ifloadreferences%
!expandafter!ifx!csname _@#1@##1!endcsname!relax%
!write16{#1 ##1 not hitherto defined - rerun saving references}%
!changedreferencestrue%
!else%
!expandafter!ifx!csname _@#1@##1!endcsname!temptoken%
!else%
!write16{#1 ##1 reference has changed - rerun saving references}%
!changedreferencestrue%
!fi%
!fi%
!else%
!expandafter!edef!csname _@#1@##1!endcsname{!temptoken}%
!edef!textoutput{!write!references{\global\def\_@#1@##1{!temptoken}}}%
!textoutput%
!fi}}%
!global!def!makecounter#1{!_@@internal@@makelabel{#1}!_@@internal@@makeref{#1}}%
!unsetcatcodes%
}
%
%
%
%
%
\def\turnintolatin#1{\ifcase #1 _\or i\or ii\or iii\or iv\or v\or vi\or vii\or viii\or ix\or x\or xi\or xii\or xiii\or xiv\or xv\or xvi\or xvii\or xviii\or xix\or xx\or xxi\or xxii\or xxiii\or xxiv\or xxv\or xxvi\fi}%
\def\alphanum#1{\ifcase #1 _\or A\or B\or C\or D\or E\or F\or G\or H\or I\or J\or K\or L\or M\or N\or O\or P\or Q\or R\or S\or T\or U\or V\or W\or X\or Y\or Z\fi}%
\newwrite\references%
\ifloadreferences{\catcode`\@=11 \catcode`\_=11 \input references.tex }%
\else{\openout\references=references.tex }%
\fi%
%
%
\newcount\headno%
\global\headno=0%
\def\headinfo{\ifinappendices\alphanum\headno\else\the\headno\fi}%
\def\nextheadno{\global\advance\headno by 1 \global\subheadno=0 \global\procno=0 \global\eqnno=0 \headinfo}%
\makecounter{head}%
%
%
\newcount\subheadno%
\global\subheadno=0%
\def\subheadinfo{\the\subheadno}%
\def\nextsubheadno{\global\advance\subheadno by 1 \global\procno=0 \subheadinfo}%
\makecounter{subhead}%
%
%
\newcount\procno%
\global\procno=0%
\def\procinfo{\subheadinfo.\the\procno}%
\def\nextprocno{\global\advance\procno by 1 \procinfo}%
\makecounter{proc}%
%
%
\newcount\figno%
\global\figno=0%
\def\figinfo{\subheadinfo.\the\figno}%
\def\nextfigno{\global\advance\figno by 1 \figinfo}%
\makecounter{fig}%
%
%
\newcount\eqnno%
\global\eqnno=0%
\def\eqninfo{\text{{\rm (\the\eqnno)}}}%
\def\nexteqnno[#1]{\global\advance\eqnno by 1 \eqninfo\hbox{\eqnlabel{#1}}}%
\makecounter{eqn}%
%
%
%
%
%
\def\gobbleeight#1#2#3#4#5#6#7#8{}%
\newcount\citationno%
\global\citationno=0%
\def\citationinfo{\the\citationno}%
\makecounter{citation}%
\newwrite\biblio%
\def\newref#1#2{%
\def\temptext{#2}%
\edef\bibliotextoutput{\expandafter\gobbleeight\meaning\temptext}%
\global\advance\citationno by 1\citationlabel{#1}%
\ifmakebiblio%
    \edef\fileoutput{\write\biblio{\noindent\hbox to 0pt{\hss$[\the\citationno]$}\hskip 0.2em\bibliotextoutput\medskip}}%
    \fileoutput%
\fi}%
\def\cite#1{%
$[\citationref{#1}]$%
\ifmakebiblio%
    \edef\fileoutput{\write\biblio{#1}}%
    \fileoutput%
\fi%
}%
%
%
%
%
\let\mypar=\par%
\edef\Pagetitle={Blank}\headline={\hfil\Pagetitle\hfil}%
\edef\Pagefooter={Blank}\footline={\hfil\Pagefooter\hfil}%
%
%
\newcount\showpagenumflag%
\global\showpagenumflag=0 %
\def\nextoddpage%
{\newpage\ifodd\pageno%
\else\global\showpagenumflag=0 %
\null\vfil\eject%
\global\showpagenumflag=1 %
\fi}%
%
%
\font\headfont=cmb12%
\def\newhead#1[#2]%
{\ifhmode\mypar\fi%
\ifnum\headno=0 \else\goodbreak\bigskip\fi%
{\headfont\noindent\nextheadno\ - #1.}\headlabel{#2}%
\nobreak\medskip}%
%
%
\def\newsubhead#1[#2]%
{\ifhmode\mypar\fi%
\ifnum\subheadno=0 \else\goodbreak\medskip\fi%
{\bf\noindent\nextsubheadno\ - #1.\ }\subheadlabel{#2}}%
%
%
\newif\ifinproclaim%
\global\inproclaimfalse%
\def\proclaim#1{%
\goodbreak\medskip
\bgroup\inproclaimtrue%
\noindent{\bf #1}%
\nobreak\medskip\sl}%
\def\noskipproclaim#1{%
\goodbreak\medskip%
\bgroup\inproclaimtrue%
\noindent{\bf #1}\nobreak\sl}%
\def\endproclaim{\mypar\egroup\nobreak\medskip\ignorespaces}%
%
%
%
\newcount\xpos\newcount\ypos
\def\makelabelgrid{%
\xpos=-5 \ypos=-5 %
\loop\ifnum\xpos<6 %
{\loop\ifnum\ypos<6 %
\def\labeltext{x}%
\ifnum\xpos=0\def\labeltext{+}\fi%
\ifnum\ypos=0\def\labeltext{+}\fi%
\placelabel[\xpos][\ypos]{\labeltext}%
\advance\ypos by 1 %
\repeat}%
\advance\xpos by 1 %
\repeat}%
\def\placelabel[#1][#2]#3{{%
\setbox10=\hbox{\raise #2cm \hbox{\hskip #1cm #3}}%
\ht10=0pt \dp10=0pt \wd10=0pt \box10}}%
%
%
%
%
\def\myitem#1{\noindent\hbox to .5cm{\hfill#1\hss}}%
%
%
%
%
%
%
%
%
%
\font\sansseriften=cmss10%
\font\sansserifseven=cmss7%
\font\sansseriffive=cmss5%
\newfam\sansseriffam%
\textfont\sansseriffam=\sansseriften%
\scriptfont\sansseriffam=\sansserifseven%
\scriptscriptfont\sansseriffam=\sansseriffive%
\def\mathsf{\fam\sansseriffam}%
%
%
%
\font\boldten=cmb10%
\font\boldseven=cmb7%
\font\boldfive=cmb5%
\newfam\mathboldfam%
\textfont\mathboldfam=\boldten%
\scriptfont\mathboldfam=\boldseven%
\scriptscriptfont\mathboldfam=\boldfive%
\def\mathbf{\fam\mathboldfam}%
%
%
%
\font\mycmmiten=cmmi10%
\font\mycmmiseven=cmmi7%
\font\mycmmifive=cmmi5%
\newfam\mycmmifam%
\textfont\mycmmifam=\mycmmiten%
\scriptfont\mycmmifam=\mycmmiseven%
\scriptscriptfont\mycmmifam=\mycmmifive%
\def\hexa#1{\ifcase #1 0\or 1\or 2\or 3\or 4\or 5\or 6\or 7\or 8\or 9\or A\or B\or C\or D\or E\or F\fi}%
\mathchardef\mathi="7\hexa\mycmmifam7B%
\mathchardef\mathj="7\hexa\mycmmifam7C%
%
%
\font\mymsbmten=msbm10 at 8pt%
\font\mymsbmseven=msbm7 at 5.6pt
\font\mymsbmfive=msbm5 at 4pt%
\newfam\mymsbmfam%
\textfont\mymsbmfam=\mymsbmten%
\scriptfont\mymsbmfam=\mymsbmseven%
\scriptscriptfont\mymsbmfam=\mymsbmfive%
\mathchardef\mybeth="7\hexa\mymsbmfam69%
\mathchardef\mygimmel="7\hexa\mymsbmfam6A%
\mathchardef\mydaleth="7\hexa\mymsbmfam6B%
%
%
%
%
\def\proof{{\noindent\bf Proof:\ }}%
\def\remark{{\noindent\bf Remark:\ }}%
\def\qed{~$\square$}%
\def\makeop#1{\global\expandafter\def\csname op#1\endcsname{{\text{#1}}}}%
\def\makeopsmall#1{\global\expandafter\def\csname op#1\endcsname{{\text{\lowercase{#1}}}}}%
%
%
%
%
%
%
\makeop{Ext}%
\makeop{Int}%
\makeop{Dist}%
\makeop{Diam}%
\makeop{Length}%
%
%
%
%
%
%
%
%
%
%
%
%
\makeop{Dim}%
\makeop{Ker}%
\makeop{Coker}%
\makeop{Tr}%
\makeop{Adj}%
\makeop{Det}%
\makeop{End}%
\makeop{Lin}%
\makeop{Symm}%
\makeop{Mult}%
%
%
\makeop{dx}%
\makeop{dy}%
\makeop{dz}%
\makeop{dt}%
\makeop{dVol}%
\makeop{dArea}%
\makeop{Supp}%
\makeop{Hess}%
\makeop{Lip}%
%
%
\makeop{Re}%
\makeop{Im}%
\makeop{Arg}%
\makeop{Log}%
\makeop{Exp}%
%
%
\makeopsmall{Cos}%
\makeopsmall{Sin}%
\makeopsmall{Tan}%
\makeopsmall{Sec}%
\makeopsmall{Cosec}%
\makeopsmall{Cot}%
\makeopsmall{ArcCos}%
\makeopsmall{ArcSin}%
\makeopsmall{ArcTan}%
\makeopsmall{ArcSec}%
\makeopsmall{ArcCosec}%
\makeopsmall{ArcCot}%
%
%
\makeopsmall{Cosh}%
\makeopsmall{Sinh}%
\makeopsmall{Tanh}%
\makeopsmall{ArcCosh}%
\makeopsmall{ArcSinh}%
\makeopsmall{ArcTanh}%
%
%
\makeop{Vol}%
\makeop{Area}%
\makeop{Riem}%
\makeop{Ric}%
\makeop{Scal}%
\makeop{Euc}%
\makeop{Imm}%
\makeop{Emb}%
%
%
\makeop{Id}%
\makeop{Ad}%
\makeop{O}%
\makeop{SO}%
\makeop{SL}%
\makeop{GL}%
\makeop{Conf}%
\makeop{Homeo}%
\makeop{Diff}%
\makeop{Isom}%
%
%
\makeop{Ind}%
\makeop{Sig}%
\makeop{Spec}%
%
%
\makeop{Conv}%
\makeop{Max}%
\makeop{Min}%
\makeop{Mod}%
\makeop{Deg}%
\makeop{loc}%
%
%
%
%
%
%
%
%
%
%
%
%
%

%% file: references.tex
\global\def\_@citation@AdlerEtAl{1}
\global\def\_@citation@BajnokEtAl{2}
\global\def\_@citation@BobenkoBook{3}
\global\def\_@citation@Bobenko{4}
\global\def\_@citation@BobenkoKuksin{5}
\global\def\_@citation@BurstallFerusPeditPinkall{6}
\global\def\_@citation@CorriganA{7}
\global\def\_@citation@CorriganEtAlA{8}
\global\def\_@citation@CorriganEtAlB{9}
\global\def\_@citation@Delius{10}
\global\def\_@citation@FadeevTakhtajan{11}
\global\def\_@citation@FokasIts{12}
\global\def\_@citation@Fokas{13}
\global\def\_@citation@FordyWood{14}
\global\def\_@citation@GilbTrud{15}
\global\def\_@citation@HauswirthKillianSchmidt{16}
\global\def\_@citation@Hitchin{17}
\global\def\_@citation@MacIntyre{18}
\global\def\_@citation@PinkallSterling{19}
\global\def\_@citation@SklyaninI{20}
\global\def\_@citation@SklyaninII{21}
\global\def\_@subhead@Introduction{1}
\global\def\_@eqn@GeneralTodaSystem{\relax \unhbox \voidb@x \hbox {{\relax \tenrm (1)}}}
\global\def\_@eqn@DefinitionOfAlphaZero{\relax \unhbox \voidb@x \hbox {{\relax \tenrm (2)}}}
\global\def\_@eqn@SinhGordon{\relax \unhbox \voidb@x \hbox {{\relax \tenrm (3)}}}
\global\def\_@eqn@LaxPair{\relax \unhbox \voidb@x \hbox {{\relax \tenrm (4)}}}
\global\def\_@eqn@SigmaMatricesI{\relax \unhbox \voidb@x \hbox {{\relax \tenrm (5)}}}
\global\def\_@eqn@KillingFieldEqnIntro{\relax \unhbox \voidb@x \hbox {{\relax \tenrm (6)}}}
\global\def\_@eqn@PolynomialKillingFieldIntro{\relax \unhbox \voidb@x \hbox {{\relax \tenrm (7)}}}
\global\def\_@eqn@DurhamBoundaryConditions{\relax \unhbox \voidb@x \hbox {{\relax \tenrm (8)}}}
\global\def\_@eqn@KMatrixIntro{\relax \unhbox \voidb@x \hbox {{\relax \tenrm (9)}}}
\global\def\_@eqn@SklyaninConditionLaxIntro{\relax \unhbox \voidb@x \hbox {{\relax \tenrm (10)}}}
\global\def\_@eqn@SklyaninConditionIntro{\relax \unhbox \voidb@x \hbox {{\relax \tenrm (11)}}}
\global\def\_@proc@MainTheorem{1.1}
\global\def\_@subhead@LesChampsDeKilling{2}
\global\def\_@eqn@GaugeTransform{\relax \unhbox \voidb@x \hbox {{\relax \tenrm (12)}}}
\global\def\_@eqn@SpecialSpacesOfLaurentSeries{\relax \unhbox \voidb@x \hbox {{\relax \tenrm (13)}}}
\global\def\_@proc@BasicAlgebraicPropertiesOfL{2.1}
\global\def\_@eqn@KillingFieldEquation{\relax \unhbox \voidb@x \hbox {{\relax \tenrm (14)}}}
\global\def\_@eqn@RI{\relax \unhbox \voidb@x \hbox {{\relax \tenrm (15)}}}
\global\def\_@eqn@RII{\relax \unhbox \voidb@x \hbox {{\relax \tenrm (16)}}}
\global\def\_@eqn@RIII{\relax \unhbox \voidb@x \hbox {{\relax \tenrm (17)}}}
\global\def\_@eqn@RIV{\relax \unhbox \voidb@x \hbox {{\relax \tenrm (18)}}}
\global\def\_@eqn@RV{\relax \unhbox \voidb@x \hbox {{\relax \tenrm (19)}}}
\global\def\_@eqn@RVI{\relax \unhbox \voidb@x \hbox {{\relax \tenrm (20)}}}
\global\def\_@proc@LaxFormulae{2.2}
\global\def\_@eqn@KFEI{\relax \unhbox \voidb@x \hbox {{\relax \tenrm (21)}}}
\global\def\_@eqn@Commutators{\relax \unhbox \voidb@x \hbox {{\relax \tenrm (22)}}}
\global\def\_@subhead@LEspaceDeChampsDeKilling{3}
\global\def\_@eqn@InitialValuesOfRecursion{\relax \unhbox \voidb@x \hbox {{\relax \tenrm (23)}}}
\global\def\_@eqn@RecursionForPsiI{\relax \unhbox \voidb@x \hbox {{\relax \tenrm (24)}}}
\global\def\_@eqn@RecursionForU{\relax \unhbox \voidb@x \hbox {{\relax \tenrm (25)}}}
\global\def\_@eqn@RecursionForPsiII{\relax \unhbox \voidb@x \hbox {{\relax \tenrm (26)}}}
\global\def\_@eqn@RecursionForTAndS{\relax \unhbox \voidb@x \hbox {{\relax \tenrm (27)}}}
\global\def\_@eqn@ExpansionForPhi{\relax \unhbox \voidb@x \hbox {{\relax \tenrm (28)}}}
\global\def\_@proc@VanishingConditions{3.1}
\global\def\_@proc@Unidimensional{3.2}
\global\def\_@subhead@LeDeterminant{4}
\global\def\_@eqn@PSGuage{\relax \unhbox \voidb@x \hbox {{\relax \tenrm (29)}}}
\global\def\_@proc@PSGuage{4.2}
\global\def\_@subhead@SklyaninsKMatrix{5}
\global\def\_@eqn@LaxPairReal{\relax \unhbox \voidb@x \hbox {{\relax \tenrm (30)}}}
\global\def\_@eqn@KMatrix{\relax \unhbox \voidb@x \hbox {{\relax \tenrm (31)}}}
\global\def\_@eqn@KMatrixRelation{\relax \unhbox \voidb@x \hbox {{\relax \tenrm (32)}}}
\global\def\_@proc@SklyaninForLaxHolds{5.1}
\global\def\_@eqn@GeneralK{\relax \unhbox \voidb@x \hbox {{\relax \tenrm (33)}}}
\global\def\_@eqn@KillingFieldII{\relax \unhbox \voidb@x \hbox {{\relax \tenrm (34)}}}
\global\def\_@eqn@XRI{\relax \unhbox \voidb@x \hbox {{\relax \tenrm (35)}}}
\global\def\_@eqn@XRII{\relax \unhbox \voidb@x \hbox {{\relax \tenrm (36)}}}
\global\def\_@eqn@XRIII{\relax \unhbox \voidb@x \hbox {{\relax \tenrm (37)}}}
\global\def\_@proc@KillingFieldsAlongBoundary{5.3}
\global\def\_@eqn@PhiTilde{\relax \unhbox \voidb@x \hbox {{\relax \tenrm (38)}}}
\global\def\_@proc@ConjugationByKPreservesKilling{5.4}
\global\def\_@eqn@KReality{\relax \unhbox \voidb@x \hbox {{\relax \tenrm (39)}}}
\global\def\_@proc@PinkalSterlingFieldIsReal{5.5}
\global\def\_@subhead@LaConditionDeBordDeRobin{6}
\global\def\_@eqn@NewExpansionForPhi{\relax \unhbox \voidb@x \hbox {{\relax \tenrm (40)}}}
\global\def\_@eqn@PSFieldGeneralGamma{\relax \unhbox \voidb@x \hbox {{\relax \tenrm (41)}}}
\global\def\_@eqn@NRI{\relax \unhbox \voidb@x \hbox {{\relax \tenrm (42)}}}
\global\def\_@eqn@NRII{\relax \unhbox \voidb@x \hbox {{\relax \tenrm (43)}}}
\global\def\_@eqn@NRIII{\relax \unhbox \voidb@x \hbox {{\relax \tenrm (44)}}}
\global\def\_@eqn@NRIV{\relax \unhbox \voidb@x \hbox {{\relax \tenrm (45)}}}
\global\def\_@eqn@NRV{\relax \unhbox \voidb@x \hbox {{\relax \tenrm (46)}}}
\global\def\_@eqn@NRVI{\relax \unhbox \voidb@x \hbox {{\relax \tenrm (47)}}}
\global\def\_@eqn@SI{\relax \unhbox \voidb@x \hbox {{\relax \tenrm (48)}}}
\global\def\_@eqn@SII{\relax \unhbox \voidb@x \hbox {{\relax \tenrm (49)}}}
\global\def\_@eqn@SIII{\relax \unhbox \voidb@x \hbox {{\relax \tenrm (50)}}}
\global\def\_@eqn@SIV{\relax \unhbox \voidb@x \hbox {{\relax \tenrm (51)}}}
\global\def\_@eqn@NSI{\relax \unhbox \voidb@x \hbox {{\relax \tenrm (52)}}}
\global\def\_@eqn@NSII{\relax \unhbox \voidb@x \hbox {{\relax \tenrm (53)}}}
\global\def\_@eqn@RobinBoundaryCondition{\relax \unhbox \voidb@x \hbox {{\relax \tenrm (54)}}}
\global\def\_@proc@BoundaryConditionsForImaginaryPart{6.4}
\global\def\_@subhead@LaSolutionEstDeTypeFini{7}
\global\def\_@proc@PreFiniteType{7.1}
\global\def\_@proc@FiniteTypeCriteria{7.2}
\global\def\_@eqn@LinearisedSinhGordon{\relax \unhbox \voidb@x \hbox {{\relax \tenrm (55)}}}
\global\def\_@proc@Liouville{7.4}
\global\def\_@subhead@DesChampsDeKillingPolynomes{8}
\global\def\_@proc@ExceptionalKillingField{8.1}
\global\def\_@head@Bibliography{A}